\documentclass[twocolumn,aps,prd,superscriptaddress,preprintnumbers,nofootinbib,showpacs]{revtex4-1}

\usepackage{amsmath}
\usepackage{graphicx}
\usepackage{subcaption}
\graphicspath{{figures/}{fig/}}
\usepackage[colorlinks=true, allcolors=blue]{hyperref}
\usepackage{slashed}
\usepackage{epstopdf}
\begin{document}

\title{Analytical two-loop amplitudes of $e^{+} e^{-} \longrightarrow \boldsymbol{J} / \boldsymbol{\psi}+\boldsymbol{\eta}_c$ at $B$ factories }

\author{Xiang Chen}
\email{xiang.chen@physik.uzh.ch}
\affiliation{Physik-Institut, Universit\"{a}t Z\"{u}rich, Winterthurerstrasse 190, CH-8057 Z\"{u}rich, Switzerland}

\author{Xin Guan}
\email{guanxin@slac.stanford.edu}
\affiliation{SLAC National Accelerator Laboratory, Stanford University, Stanford, CA 94039, USA}

\author{Chuan-Qi He}
\email{legend\_he@m.scnu.edu.cn}
\affiliation{State Key Laboratory of Nuclear Physics and Technology, Institute of Quantum Matter, South China Normal University, Guangzhou 510006, China}
\affiliation{Guangdong Basic Research Center of Excellence for Structure and Fundamental Interactions of Matter, Guangdong Provincial Key Laboratory of Nuclear Science, Guangzhou 510006, China}
\affiliation{Department of Physics and Astronomy, University of California, Los Angeles, CA 90095, USA}

\author{Yan-Qing Ma}
\email{yqma@pku.edu.cn}
\affiliation{School of Physics, Peking University, Beijing 100871, China}
\affiliation{Center for High Energy Physics, Peking University, Beijing 100871, China}

\author{Jian Wang}
\email{j.wang@sdu.edu.cn}
\affiliation{School of Physics, Shandong University, Jinan, Shandong 250100, China}
\affiliation{Center for High Energy Physics, Peking University, Beijing 100871, China}

\author{Da-Jiang Zhang}
\email{zhangdajiang@mail.sdu.edu.cn}
\affiliation{School of Physics, Shandong University, Jinan, Shandong 250100, China}

\begin{abstract}
 In double charmonium production, a long-standing challenge is that the theoretical predictions are not consistent with the measurements at B factories. Within the NRQCD framework, the next-to-leading order (NLO) calculation has proved its power to cut down the discrepancy between theory and experiments. To further clarify this puzzle, we have performed the next-to-next-to-leading order (NNLO) calculation. The amplitude is obtained as an analytical asymptotic expansion in the ratio of the squared charm-quark mass over the squared center-of-mass energy, $m_c^2/s$.
 We investigate the origin of the leading logarithms by performing a region analysis, revealing the intricate factorization structure in this process.
 We provide numerical predictions on the total cross sections of $J/\psi+\eta_c$ production, which agree with the experimental results. 
 Extension of our computation to $\Upsilon+\eta_b$ production is also discussed.
 
\end{abstract}

\maketitle

\section{Introduction}
Back in the beginning of this century, exclusive double charmonium production  in electron-positron annihilation at B factories has drawn great theoretical interest \cite{PhysRevLett.89.142001,Belle:2004abn,BaBar:2005nic}. Since these sorts of hard exclusive reactions serve as a novel and fertile ground for exploring the interplay between perturbative and non-perturbative aspects in the heavy flavor physics of QCD, many studies of double quarkonium production including $e^{+} e^{-} \rightarrow J / \psi+H$ have been performed by several research groups \cite{PhysRevD.67.054007,PhysRevD.70.074007,Bondar:2004sv,PhysRevLett.52.1080}, where $H$ denotes a C-even charmonium that recoils against $J / \psi$. Recently, the Belle II data-taking is in progress \cite{Belle:2023gln}, which allows to probe the heavy flavor physics with unprecedented precision \cite{Belle-II:2018jsg}. Future high-luminosity lepton colliders such as the ILC, CEPC, and FCC-ee are expected to provide complementary opportunities to study double charmonium production~\cite{CEPCStudyGroup:2018ghi,Duran:2025ogk,FCC:2018evy}. Thus, a reliable theoretical prediction is highly demanded.

Pioneering theoretical effort, including leading order (LO) NRQCD, light-cone or general pQCD approaches \cite{PhysRevD.67.054007,PhysRevD.70.074007,Bondar:2004sv,PhysRevLett.52.1080}, exhibited unsatisfactory large uncertainties and discrepancies with experimental data.
Especially, the LO NRQCD predictions \cite{PhysRevD.67.054007} of $e^{+} e^{-} \rightarrow J / \psi+\eta_c$ is much smaller than the measured cross section~\cite{PhysRevLett.89.142001,BaBar:2005nic}. Here we list the experimental results:
\begin{equation}\label{belle}
\begin{aligned}
& \sigma(J / \psi+\eta_c(1 S)) \mathcal{B}(2)=25.6 \pm 2.8 \pm 3.4 \mathrm{fb}\,, \\
& \sigma(J / \psi+\eta_c(2 S)) \mathcal{B}(2)=16.5 \pm 3.0 \pm 2.4 \mathrm{fb}\,,
\end{aligned}
\end{equation}
from Belle, and
\begin{equation}\label{babar}
\begin{aligned}
& \sigma\left(J / \psi+\eta_c(1 S)\right) \mathcal{B}(2)=17.6 \pm 2.8_{-2.1}^{+1.5} \mathrm{fb}\,, \\
& \sigma\left(J / \psi+\eta_c(2 S)\right) \mathcal{B}(2)=16.4 \pm 3.7_{-3.0}^{+2.4} \mathrm{fb}\,,
\end{aligned}
\end{equation}
from BaBar, where $\mathcal{B}(n)$ denotes the branching fraction for $\eta_c$ decaying into more than $n$ charged tracks.
In principle, the NRQCD theory \cite{Bodwin:1994jh} provides an effective framework that allows us to systematically deal with the higher-order corrections, which may be helpful to understand this discrepancy.

The next-to-leading order (NLO) perturbative calculations within the NRQCD framework were computed in Ref.~\cite{Zhang:2005cha}, which, for the first time, reaches the lower uncertainty bound of the experimental measurements. This computation demonstrates the  importance of high-order corrections in the $e^{+}e^{-} \rightarrow J/\psi + \eta_c$ process. The NLO calculation was later verified independently by \cite{Gong:2007db}.  
Subsequent developments addressed the relativistic corrections for the process up to $\mathcal{O}(v^2)$ \cite{Braaten:2002fi,He:2007te}, where $v$ represents the characteristic velocity of the constituent charm quarks in the quarkonium rest frame.  While the non-perturbative NRQCD matrix elements inherently carry significant theoretical uncertainties, comprehensive studies demonstrated that combining both NLO perturbative corrections and partial resummation of relativistic effects \cite{PhysRevD.77.094018} could yield predictions that were broadly consistent with B-factory measurements, albeit with substantial residual uncertainties.  Further refinement came with the investigation of joint perturbative and relativistic corrections at $\mathcal{O}(\alpha_s v^2)$, which were found to provide modest but meaningful enhancements to the NRQCD theoretical predictions \cite{PhysRevD.85.114018}.

%A crucial progress resolving the large discrepancy is brought by the next-to-leading order perturbative calculation, where $\mathcal{O}\left(\alpha_s\right)$ contribution is included \cite{Zhang:2005cha}.Within years, researchers addressed the relativistic correction to $e^{+} e^{-} \rightarrow J / \psi+\eta_c$ to $\mathcal{O}\left(v^2\right)$, where $v$ denotes the velocity of charm quarks \cite{Braaten:2002fi,He:2007te}.Although the various NRQCD matrix elements possess inevitable uncertainty, it was believed that, including both NLO perturbative and (a partial resummation of) relativistic corrections\cite{PhysRevD.77.094018}, one may achieve reasonable agreement, albeit with large uncertainties, between the NRQCD prediction and B factory data.Later the joint perturbative and relativistic $\mathcal{O}\left(\alpha_s v^2\right)$ correction was also investigated, which was found to modestly enhance the existing NRQCD predictions \cite{PhysRevD.85.114018}.

The next-to-next-to-leading order (NNLO) radiative corrections were numerically computed  in Ref. \cite{Feng:2019zmt} and further confirmed in Ref. \cite{Huang:2022dfw}, which were found to be substantial. Although the inclusion of both radiative and relativistic effects brings the theoretical prediction into consistent with experimental measurements, significant uncertainties persist, primarily due to the choice of renormalization scale. It is therefore essential to investigate the origin of the poor perturbative convergence in this process. At NLO, the dominant contribution comes from the leading logarithm, as analyzed in Refs. \cite{Jia:2010fw,Bodwin:2014dqa}. However, the significance of logarithmic contributions at NNLO remains unclear.

In this paper, we provide a detailed computation of the NNLO corrections of the process $e^{+} e^{-} \rightarrow J / \psi+\eta_c$, utilizing state-of-the-art  techniques. Our results are expressed as an asymptotic expansion in the variable $r=m_c^2/s$, enabling us to extract all logarithmic terms of $r$. The concise leading logarithmic terms are shown explicitly in the main text, while more extensive subleading terms are provided in an attached supplementary file. Furthermore, we explore the origin and structure of the leading logarithmic contributions and summarize the resulting theoretical insights.
Using our analytical results, we provide physical predictions for the $J / \psi+\eta_c$ production cross section at various center-of-mass energies. As a nontrivial byproduct, our formalism also allows for a convenient prediction of the $\Upsilon+\eta_b$ cross section, which may be tested in future experiments. We additionally investigate the impact of input parameters and scale choices on the phenomenological outcomes and assess the convergence behavior of the perturbative expansion. The paper concludes with a summary of our main results.

\section{Framework and Methods}
We consider the two-loop QCD correction to the following process
\begin{equation}
    \label{equ:process}
    e^+(k_1)+e^-(k_2)\rightarrow J / \psi(P_1)+\eta_c(P_2).
\end{equation}
At the lowest order in velocity expansion within NRQCD, the scalar products of external momenta satisfy
\begin{align}
    k_1^2 = k_2^2 = 0,  \quad
    P_1^2 = P_2^2 = 4 m_c^2, \quad
   (k_1 + k_2)^2 = s,
\end{align}
where $\sqrt{s}$ is the  center-of-mass energy and $m_{c}$ denotes the charm-quark mass. 
It has been proved that the cross section of $e^{+} e^{-} \rightarrow J / \psi+\eta_c$  can be factorized into the leptonic and hadronic parts in Ref. \cite{Sun:2021tma}. 

\begin{figure}[htb]
	\centering
    \begin{minipage}[b]{.49\linewidth}
        \centering
        \includegraphics[width=1.05\linewidth]{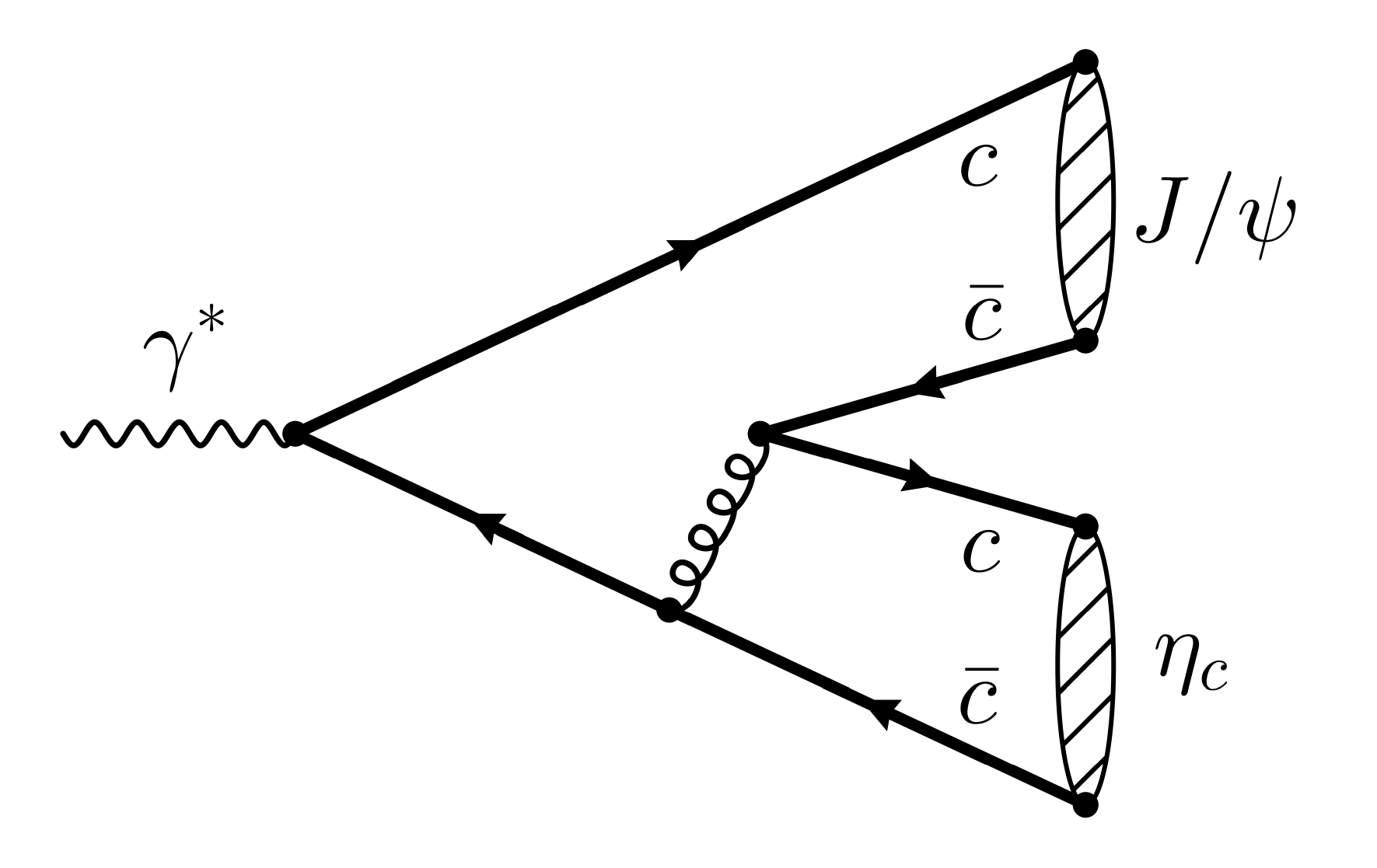}
    \subcaption{}
    \end{minipage}
    \begin{minipage}[b]{.49\linewidth}
        \centering
        \includegraphics[width=1.05\linewidth]{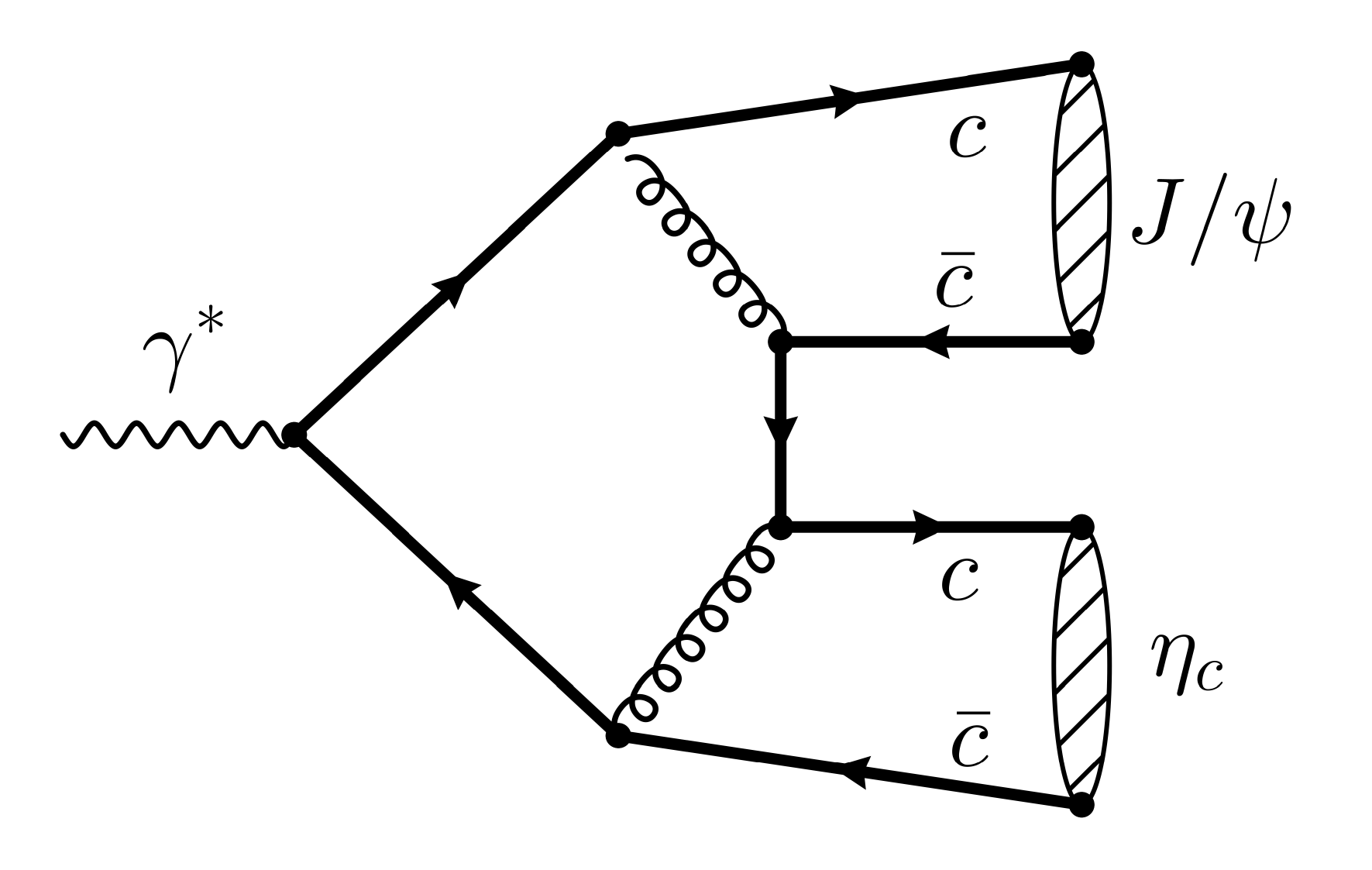}
    \subcaption{}
    \end{minipage}
    \caption{Selected tree-level and one-loop diagrams for $\gamma^* \rightarrow J/\psi+\eta_{c}$.}\label{fig:diagloop1}
\end{figure}

It is convenient to define the amplitudes of the process $\gamma^{\ast} \rightarrow J / \psi+\eta_c$ as 
\begin{equation}\label{eq:ScatteringME}
\begin{aligned}
&\mathcal{A}^{\mu}(s)=\left\langle J / \psi(P_1, \lambda)+\eta_c(P_2)\left|J^\mu\right| 0\right\rangle \\
=&\sum_{i=0} h^{(i)}(r) \frac{e}{192 \pi m_c^5}g_s^{2i+2}\epsilon^{\sigma \mu P_{1} P_{2}}\varepsilon_\sigma^*(\lambda)\\
&\times \sqrt{M_{J/\psi}M_{\eta_c}}R_{J/\psi}(0)R_{\eta_c}(0), 
\end{aligned}
\end{equation}
%\cqcom{[Because of definition in (5), $h(r)$ is purely imaginary.Maybe we should use $i\mathcal{A}(s)$=... .]}
with $J^{\mu}$ being the electromagnetic current and $h^{(i)}(r)$ being dimensionless expansion terms in $\alpha_s$. The quarkonium mass can be taken as $M_H^2 \approx 4 m_c^2$ and $R_H(0)$ denotes the wave function at the origin.
Typical tree-level and one-loop Feynman diagrams are shown in Fig. \ref{fig:diagloop1}.

With the amplitudes at hand, the total cross section for $e^{+} e^{-} \rightarrow J / \psi+\eta_c$ can be written as
\begin{equation}
\sigma=\frac{\alpha}{12 s^2}\sqrt{\frac{s-16m_c^2}{s}} \mathcal{A}^{\mu}\mathcal{A}_{\mu}^{*}.
\end{equation}
At the LO, we have
\begin{equation}\label{eq:LO}
h^{(0)}(r) =-i \frac{512 \left(\mathrm{N}_c^2-1\right)}{3} r^2.
\end{equation}
In the following, we will compute $h^{(1)}(r)$ and $h^{(2)}(r)$. 

\begin{figure}[htb]
	\centering
    \begin{minipage}[b]{.49\linewidth}
        \centering
        \includegraphics[width=1.05\linewidth]{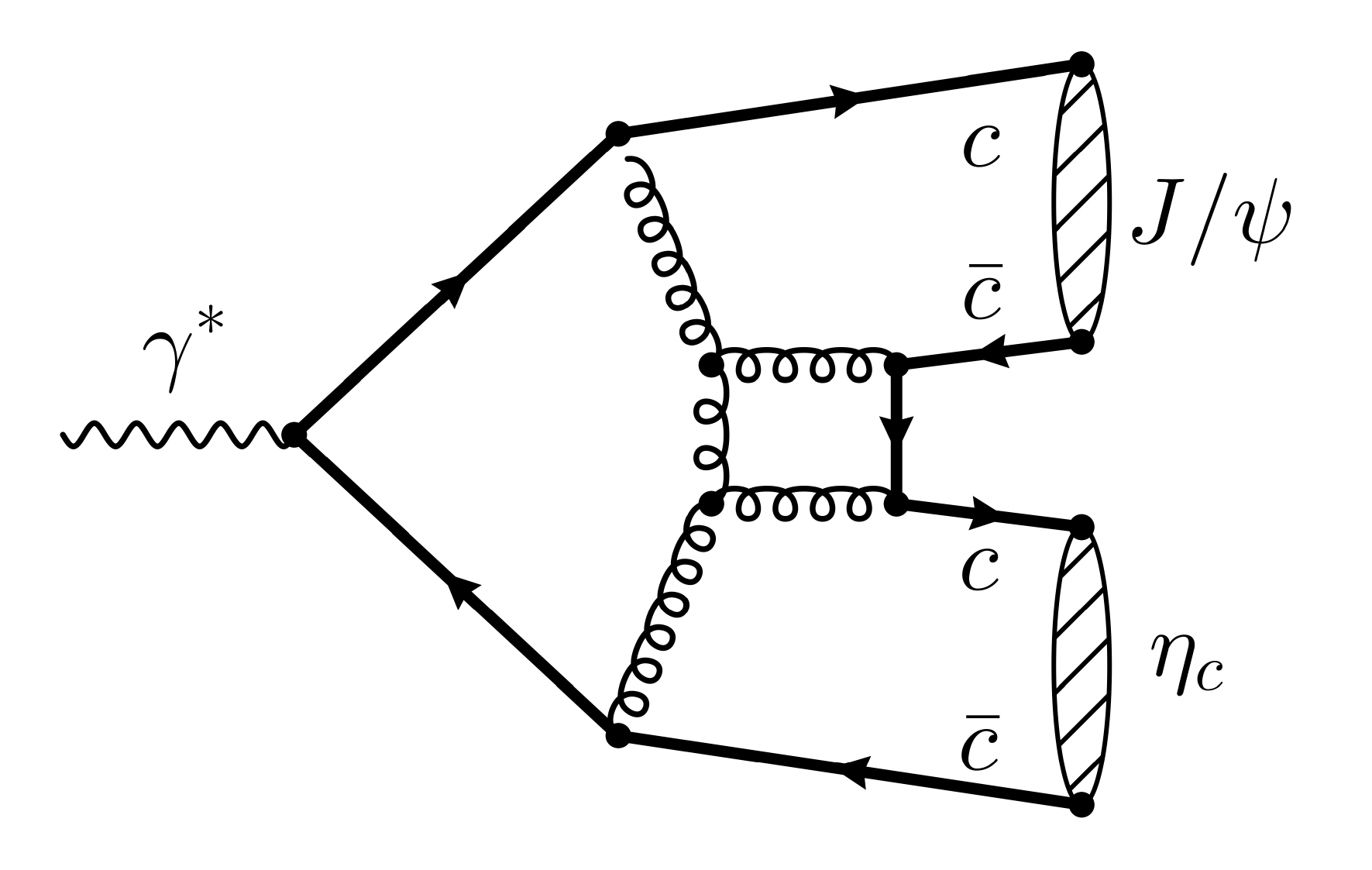}
    \subcaption{}
    \end{minipage}
    \begin{minipage}[b]{.49\linewidth}
        \centering
        \includegraphics[width=1.05\linewidth]{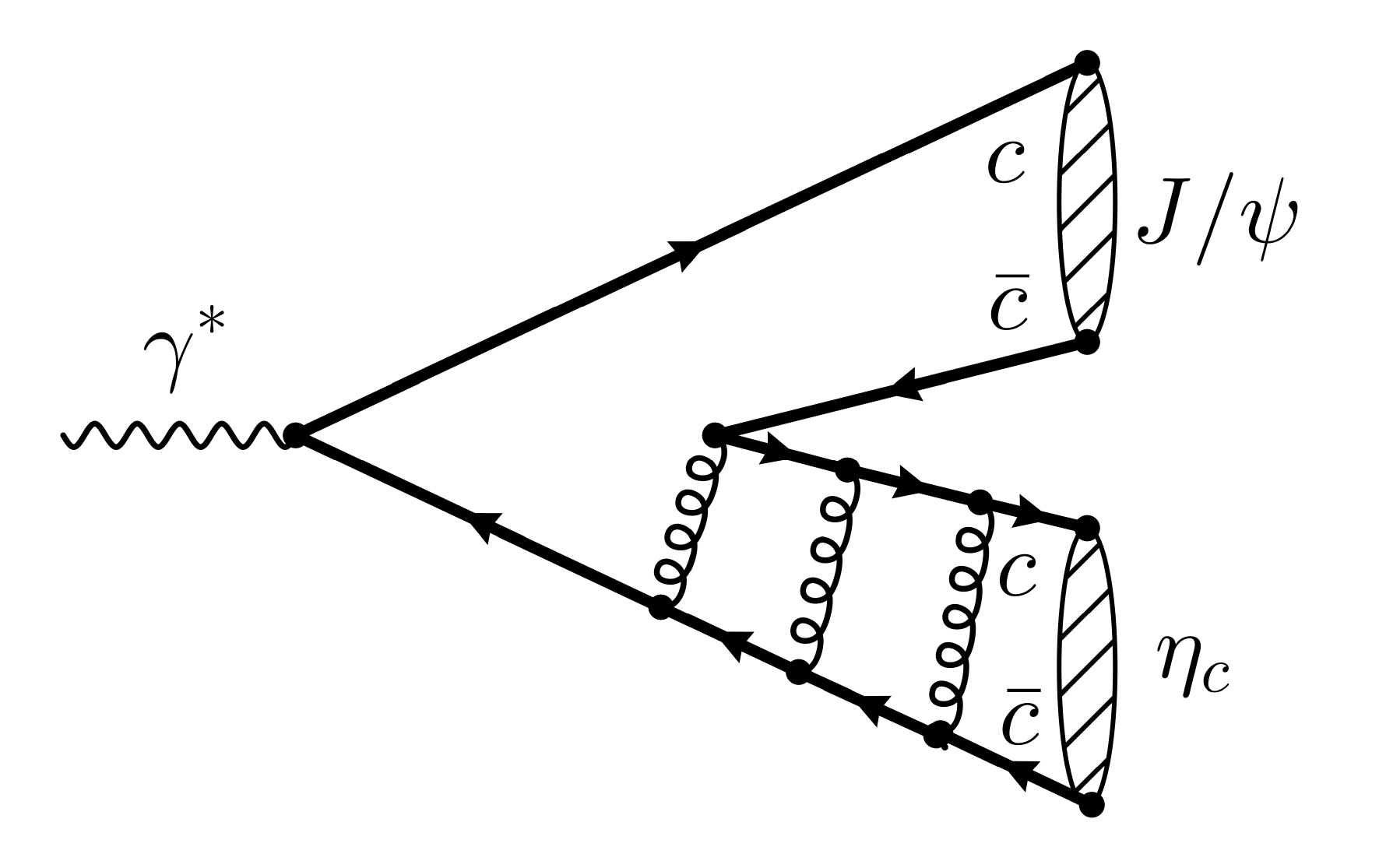}
    \subcaption{}
    \end{minipage}
    \begin{minipage}[b]{.49\linewidth}
        \centering
        \includegraphics[width=1.05\linewidth]{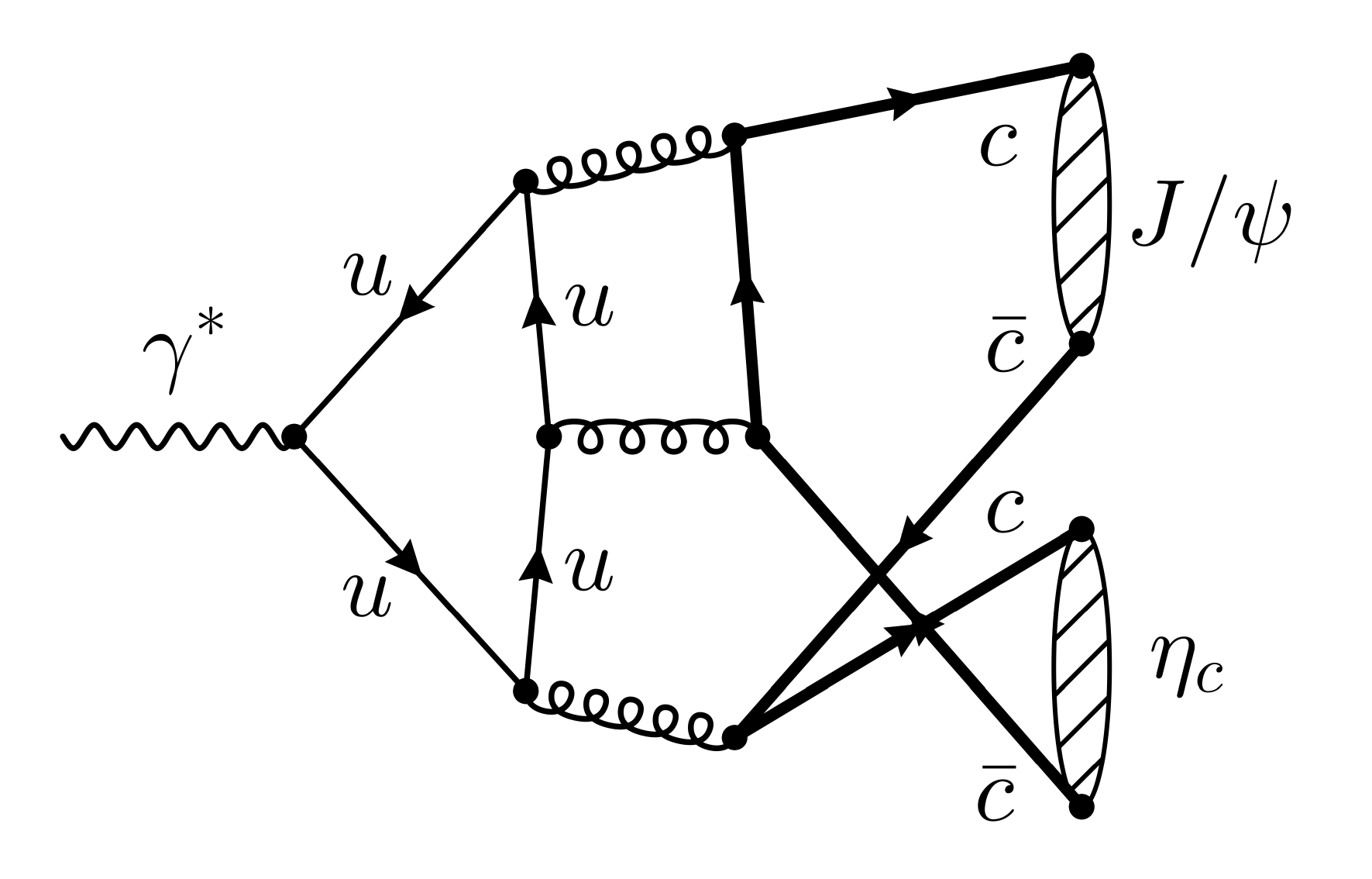}
    \subcaption{}
    \end{minipage}
    \begin{minipage}[b]{.49\linewidth}
        \centering
        \includegraphics[width=1.05\linewidth]{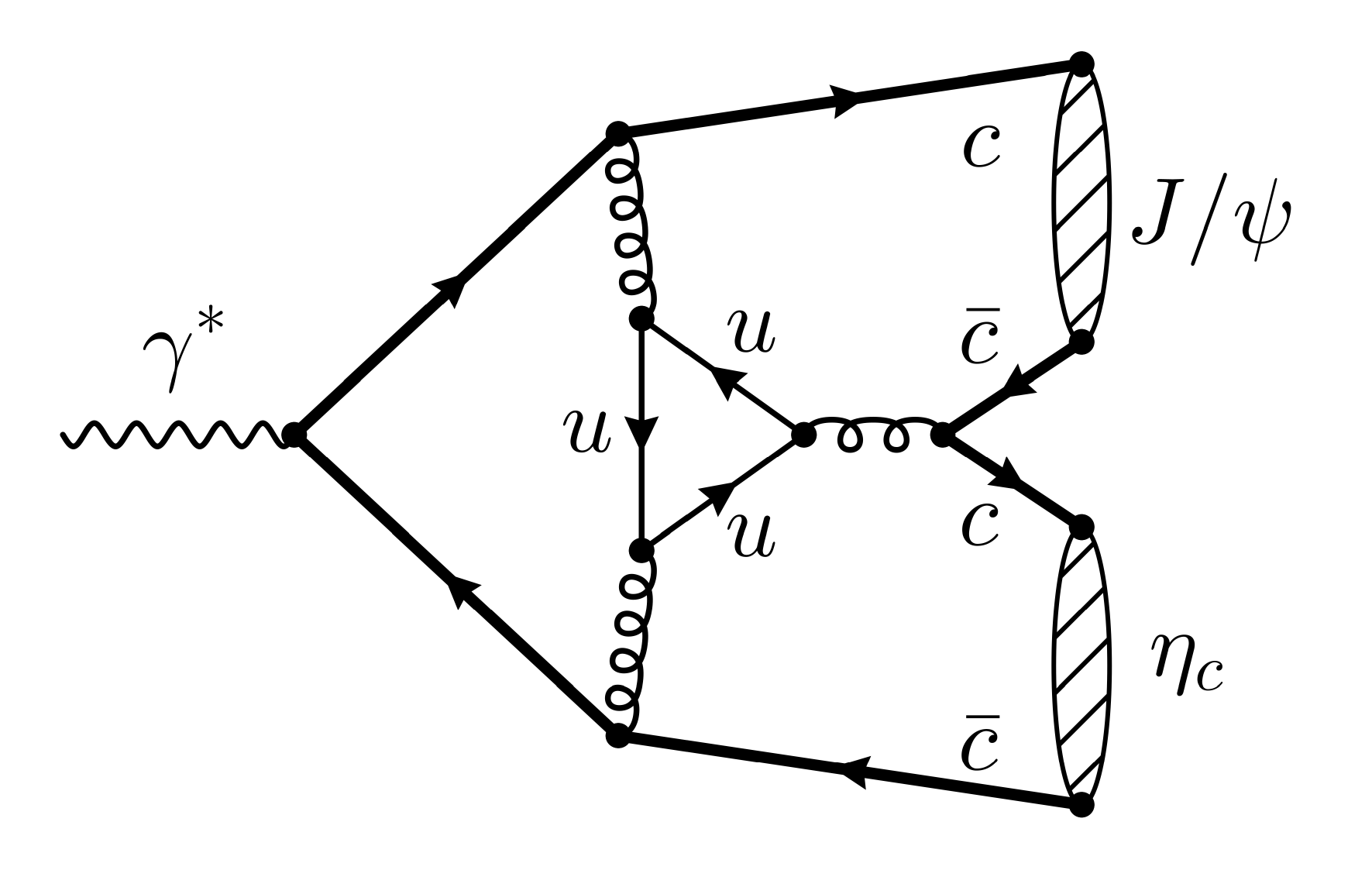}
    \subcaption{}
    \end{minipage}
    \caption{Selected  two-loop diagrams for $\gamma^* \rightarrow J/\psi+\eta_{c}$.}
    \label{fig:diagloop2}
\end{figure}

We generate the Feynman diagrams for the process with {\tt FeynArts}~\cite{Nogueira:1991ex} and {\tt QGRAF}~\cite{Hahn:2000kx}, with some representative two-loop diagrams shown in Fig.~\ref{fig:diagloop2}. Our calculation employs the dimensional regularization with $D = 4 -2 \epsilon$. The amplitudes are computed using the {\tt {CalcLoop}} package \cite{calcloop}, which performs the Dirac algebra, Lorentz contractions, as well as $SU(N_c)$ color algebra operations. Through tensor decomposition,  we express the amplitudes as linear combinations of scalar Feynman integrals, which are classified into 167 distinct integral families. For each family, we utilize the packages {\tt Blade}~\cite{Guan:2024byi}, {\tt FiniteFlow}~\cite{Peraro:2016wsq,Peraro:2019svx} and  {\tt LiteRed}~\cite{Lee:2012cn} to perform the integration-by-part (IBP)~\cite{Tkachov1981,Chetyrkin1981,Laporta:2001dd} reduction to express these integrals as linear combinations of a set of so-called master integrals. In our computation, we assign $\epsilon$ small numerical values and fit the $\epsilon$ dependence at the final stage as proposed in Refs.~\cite{Liu:2022mfb,Liu:2022chg}, which avoids the manipulation of $\epsilon$ in the intermediate step and reduces the computational time. 

To evaluate the master integrals, we use the method of numeric differential equations~\cite{Kotikov:1990kg}. We construct the differential equations with respect to the dimensionless variable $r=m_c^2/s$. The boundary conditions at $r=1/100$ are obtained using the auxiliary mass flow method~\cite{Liu:2017jxz,Liu:2021wks, Liu:2022mfb}, implemented in {\tt AMFlow}~\cite{Liu:2022chg}. We note that spurious or non-physical singularities can cause numerical instability, which can be resolved by increasing the working precision in our numerical computation. Feynman integrals exhibit the following power series expansion structure~\cite{Caffo:2008aw, Czakon:2008zk,Lee:2014ioa,Moriello:2019yhu, Hidding:2020ytt, Armadillo:2022ugh}:
\begin{equation}\label{eq:asy0}
\mathcal{I}(r)=\sum_{ k, n} c_{k, n}(\epsilon) \ln ^k(r) r^n  .
\end{equation}
We utilize the package {\tt DESolver} implemented in {\tt AMFlow} to solve the differential equation as a power series. 

The precision of the resulting asymptotic expansion coefficients $c_{k,n}$ depends on both the accuracy of the boundary conditions and the truncation order of $r$.  We truncate the expansion to at most $O(r^{2000})$ to ensure that the leading expansion coefficients have at least 600 digits of precision.  The high precision results enable us to reconstruct the analytic form of the leading coefficients using the PSLQ algorithm \cite{1992A}. The basis we use in this work for the PSLQ reconstruction of finite terms is the combination of three basic elements $\mathrm{Li}_m(-2),\mathrm{Li}_m(-1),\mathrm{Li}_m(2)$ for $m \leq 4$. With the analytic form of the leading coefficients as boundary conditions, we can then solve the differential equations again to obtain the analytic form for all other coefficients. Finally, we express the amplitudes as a series expansion in $r$ with fully analytical coefficients, enabling exact symbolic manipulation and further theoretical analysis.

\section{Renormalization and factorization}
The amplitude, before factorization matching, can be written as
\begin{equation}
\tilde{\mathcal{A}}^{\mu}(\alpha_s, m_Q)=Z_{2, Q}^{-2}\left(\mathcal{A}_{0}^{\mu(0)}+\mathcal{A}_{0}^{\mu(1)}+\mathcal{A}_{0}^{\mu(2)}\right),
\end{equation}
where $Z_{2, Q}$ is the field renormalization constant for the heavy quarks in the final state, and $\mathcal{A}_{0}^{\mu(i)}$ are  amputated bare amplitudes which in general depend on bare $\alpha_{s, 0}$ and $ m_{Q, 0}$. The bare and renormalized quantities are related as
\begin{equation}
m_{Q, 0}=Z_{m, Q} m_Q, ~~
\alpha_{s, 0}=(4 \pi e^{-\gamma_E})^{-\epsilon} \mu_R^{2 \epsilon} Z_{\alpha_s}^{\overline{\mathrm{MS}}} \alpha_s(\mu_R),   
\end{equation}
%\begin{equation}
%\begin{aligned}
%&m_{Q, 0}=Z_{m, Q} m_Q, \\
%&\alpha_{s, 0}=(4 \pi e^{-\gamma_E})^{-\epsilon} \mu_R^{2 \epsilon} Z_{\alpha_s}^{\overline{\mathrm{MS}}} \alpha_s(\mu_R),   
%\end{aligned}
%\end{equation}
where $\mu_R$ is the renormalization scale. The renormalization constants can be found in Refs.~\cite{Czakon:2007wk,Czakon:2008zk,Barnreuther:2013qvf} and references therein.

After performing the UV renormalization, we are left with a single $\epsilon$ pole in the amplitude, which satisfies exactly the following factorization relation
\begin{equation}
\tilde{\mathcal{A}}^{\mu}= -\frac{1}{2\epsilon}\frac{\alpha_s^2}{\pi^2}(\gamma_{J / \psi}+\gamma_{\eta_c} )\mathcal{A}^{\mu(0)}+\mathcal{O}(\epsilon^0) ,
%\tilde{\mathcal{A}}^{\mu}= -\frac{\alpha_s^2}{2\epsilon}(\gamma_{J / \psi}+\gamma_{\eta_c} )\mathcal{A}^{\mu(0)}+\mathcal{O}(\epsilon^0) ,
\end{equation}
where the anomalous dimensions are defined as \cite{Czarnecki:2001zc,Kniehl:2006qw,PhysRevLett.80.2535,PhysRevLett.80.2531}
\begin{equation}
\begin{aligned}
\gamma_{J / \psi} & =-\frac{\pi^2}{12} C_F(2 C_F+3 C_A), \\
\gamma_{\eta_c} & =-\frac{\pi^2}{4} C_F(2 C_F+C_A).
\end{aligned}
\end{equation}
The exact subtraction of the poles from the amplitudes provides a robust examination of our calculation. The finite amplitude after factorization matching reads: 
\begin{equation}
\bar{\mathcal{A}}_{\text{finite}}^{\mu(2)}=\tilde{\mathcal{A}}^{\mu(2)}+ \frac{1}{2\epsilon}\frac{\alpha_s^2}{\pi^2}(\gamma_{J / \psi}+\gamma_{\eta_c} )\mathcal{A}^{\mu(0)} \left(\frac{\mu_{\Lambda}^2 }{\mu_R^2 }\right)^{-2 \epsilon}
%\bar{\mathcal{A}}_{\text{finite}}^{\mu(2)}=\tilde{\mathcal{A}}^{\mu(2)}- \frac{\alpha_s^2}{2 \epsilon} \tilde{\mathcal{A}}^{\mu(0)} \left(\frac{4 C_F^2}{3}+C_F C_A\right) \left(\frac{\mu_{\Lambda}^2 e^{\gamma_E}}{\mu_R^2 4 \pi}\right)^{-2 \epsilon}
\end{equation}
with $\mu_{\Lambda}$ being the factorization scale. Now, we have obtained the renormalized two-loop amplitude $\bar{\mathcal{A}}_{\text{finite}}^{\mu(2)}$, which is free of divergences. 
Unless otherwise specified, we simply use $\mathcal{A}^{\mu(i)}$ to indicate the finite amplitude in the following text.
The corresponding $h^{(i)}(r)$ functions likewise have no divergences.
%From the definition in Eq.~\eqref{eq:ScatteringME}, we only need to provide the values of $h^{(i)}(r)$. 

%\emph{Analytical Results.}---
\section{Analytical Results}

We express $h^{(i)}(r)$ explicitly as following:
\begin{equation}\label{eq:JEExpand}
\begin{aligned}
h^{(0)}(r)&= r^2 L_{(2,0)}^{(0)},  \\
h^{(1)}(r)&= r^2 \left[L_{(2,2)}^{(1)} \;  \ln^2(r) +L_{(2,1)}^{(1)} \;  \ln(r)  +L_{(2,0)}^{(1)}\right] \\
&+\mathcal{O}(r^3), \\
h^{(2)}(r)&=r^2 \left[L_{(2,4)}^{(2)}  \;  \ln^4(r) +L_{(2,3)}^{(2)} \; \ln^3(r) \right. \\
&\left. + L_{(2,2)}^{(2)} \; \ln^2(r) +L_{(2,1)}^{(2)} \; \ln(r) +L_{(2,0)}^{(2)} \right] \\
&+r^{5/2} L_{(5/2,0)}^{(2)}+\mathcal{O}(r^3),
\end{aligned}
\end{equation}
where higher-power terms in $r$ expansion are ignored since they are irrelevant for actual phenomenological study.

The LO result is given in Eq.~\eqref{eq:LO}. The logarithmic terms at NLO read
\begin{equation}
\begin{aligned}
L_{(2,2)}^{(1)}=&-\frac{4 i(\mathrm{N}_c^2-1)(7 \mathrm{N}_c^2-11) }{3 \pi^2 \mathrm{N}_c},
\end{aligned}
\end{equation}
\begin{equation}
\begin{aligned}
L_{(2,1)}^{(1)} =&  N_{lf}\frac{64 i (\mathrm{N}_c^2-1) }{9 \pi^2}+\frac{8(\mathrm{N}_c^2-1)(7 \mathrm{N}_c^2-11) }{3 \pi \mathrm{N}_c}  \\
& -\frac{4 i(\mathrm{N}_c^2-1) }{9 \pi^2 \mathrm{N}_c}\left[(31+138 \ln 2) \mathrm{N}_c^2 \right.\\
& \left.-28 \mathrm{N}_c-258 \ln 2+57\right]  ,
\end{aligned}
\end{equation}
where $N_{lf}$ denotes the number of light fermions.
These results are found to be consistent with Refs.~\cite{Jia:2010fw,Gong:2007db,Bodwin:2014dqa}.
The leading and next-to-leading logarithmic terms at NNLO read
\begin{equation}\label{eq:log4t}
L_{(2,4)}^{(2)} =-\frac{i(\mathrm{N}_c^2-1)^2(7 \mathrm{N}_c^2-19) }{144 \pi^4 \mathrm{N}_c^2},
\end{equation}
\begin{equation}\label{eq:log3t}
\begin{aligned}
L_{(2,3)}^{(2)} = &N_{lf}\frac{ 2 i (\mathrm{N}_c^2-1)(7 \mathrm{N}_c^2-11)  }{27 \pi^4 \mathrm{N}_c} \\
&+ \frac{(\mathrm{N}_c^2-1)^2\left(7 \mathrm{N}_c^2-19\right) }{36 \pi^3 \mathrm{N}_c^2} \\
&-\frac{i\left(\mathrm{N}_c^2-1\right)}{54 \pi^4 \mathrm{N}_c^2}\left[\mathrm{N}_c^4(79+87 \ln 2)-31 \mathrm{N}_c^3 \right.\\
&\left.-2 \mathrm{N}_c^2(40+141 \ln 2) +\mathrm{N}_c(59-24 \ln 2)\right.\\
&\left.-99+189 \ln 2\right].  % \bm{t}^2 ln^3(\bm{t}).
\end{aligned}
\end{equation}
Besides, the asymptotic expansion at two-loop order has an unusual non-integer power term $r^{5/2}$, with its coefficient given by
\begin{equation}\label{eq:2p5}
L_{(5/2,0)}^{(2)} =-\frac{8i(\mathrm{N}_c^2-1)(7 \mathrm{N}_c^2-2) }{9 \pi^2 \mathrm{N}_c}.
\end{equation}
Other coefficients are too lengthy to be given here, and they are provided in the supplementary materials.

%Note: "logfig.pdf" is for the contribution of ln(s) which do not include the ln(m^2) and leave them to the regular term.
%%%%%%%%%%%%%%%%%%
\begin{figure}[htb]
    \begin{minipage}[b]{1\linewidth}
        \centering
        \includegraphics[width=1\linewidth]{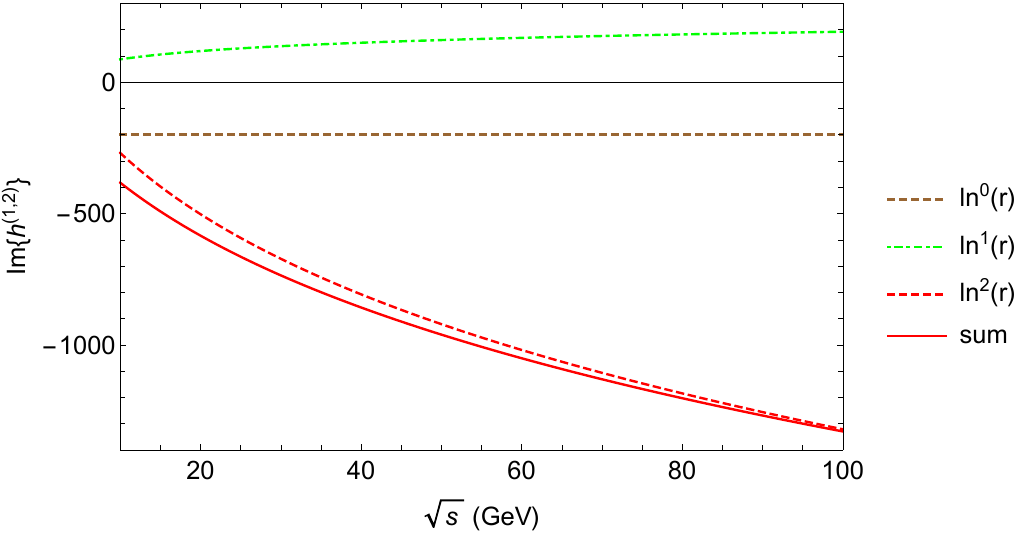}
    \subcaption{NLO amplitudes}\label{fig:nlolog}
    \end{minipage}
    \begin{minipage}[b]{1\linewidth}
        \centering
        \includegraphics[width=1\linewidth]{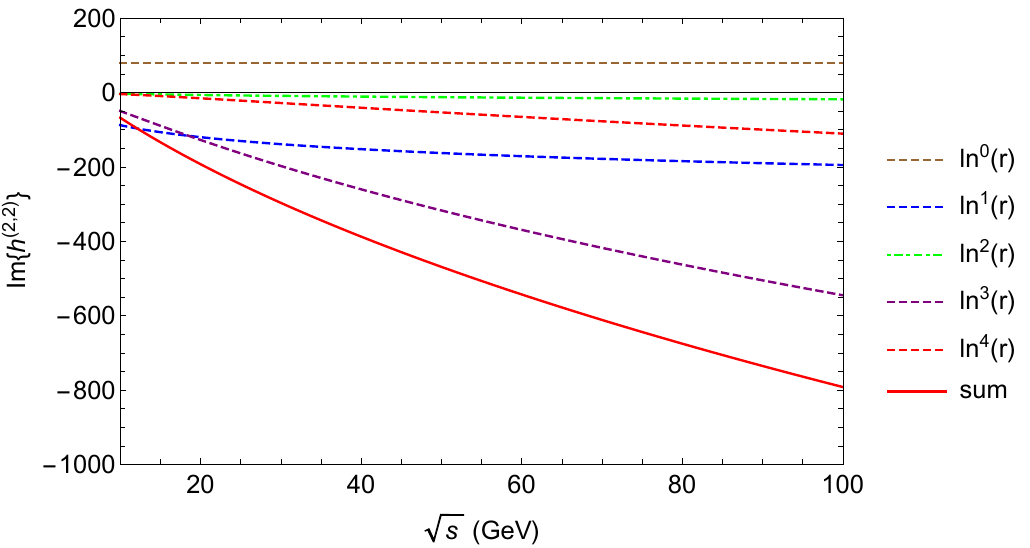}
    \subcaption{NNLO amplitudes}\label{fig:nnlolog}
    \end{minipage}
    \caption{The contribution of each logarithmic term $\ln^{i}(r)$ to the $\gamma^* \rightarrow J/\psi+\eta_{c}$ amplitudes at different energies, with $\mu_R=\sqrt{s}$ and $\mu_\Lambda=m_c=1.5~$GeV.}\label{fig:LogSignifcance}
\end{figure}
%%%%%%%%%%%%%%%%%%

We define $h^{(i,2)}(r)$ as the coefficients at leading power, given by
\begin{equation}\label{eq:JEExpand}
\begin{aligned}
h^{(0,2)}(r)&= L_{(2,0)}^{(0)} ,\\
h^{(1,2)}(r)&= L_{(2,2)}^{(1)} \;  \ln^2(r) +L_{(2,1)}^{(1)} \;  \ln(r)  +L_{(2,0)}^{(1)}, \\
h^{(2,2)}(r)&=L_{(2,4)}^{(2)}  \;  \ln^4(r) +L_{(2,3)}^{(2)} \; \ln^3(r)\\
&+ L_{(2,2)}^{(2)} \; \ln^2(r) +L_{(2,1)}^{(2)} \; \ln(r) +L_{(2,0)}^{(2)} .
\end{aligned}
\end{equation}
By setting $\mu_R=\sqrt{s}$, $\mu_\Lambda=m_c$ and $N_{lf}=3$,
we have the numerical results
\begin{equation}
\begin{aligned}
\mathrm{Im}\{h^{(0,2)}\}=&-1365.33, \\
\mathrm{Im}\{h^{(1,2)}\}=&-18.733 \ln ^2(r)-22.823 \ln (r)-199.927, \\
\mathrm{Im}\{h^{(2,2)}\}=&-0.02231 \ln ^4(r)+0.9202 \ln ^3(r) \\
&-0.2663 \ln ^2(r)+23.26 \ln (r)+78.63\,, 
\end{aligned}
\end{equation}
where only the imaginary part is given. %\yqcom{since real part does not contribute to cross section????}
With the above results, we can analyze the relative contribution from each logarithmic term at different energy points. As shown in Fig.~\ref{fig:LogSignifcance}, the $\ln^2(r)$ term dominates the NLO results. At NNLO, the $\ln^3(r)$ term dominates across most of the energy region of phenomenological  interest, while the $\ln^4(r)$ term also contributes substantially at relatively higher energies. The observation implies that resummation of both the leading and next-to-leading logarithmic terms to all orders of $\alpha_s$ should be necessary.

\section{The structure of large logarithms}
The large logarithms shown in the last section not only give a major contribution to the cross section,
but also possess intriguing features that deserve detailed exploration.
The leading logarithms at NLO have been analysed in Ref. \cite{Bodwin:2014dqa}.
It was found that the double logarithms originate solely from the soft fermion contribution while those from soft gluon contribution cancel out.
This kind of logarithm has received attention recently and has been discussed in $H\to \gamma\gamma$ \cite{Liu:2019oav,Wang:2019mym,Hou:2025ovb}, $Hgg$ form factor \cite{Liu:2018czl,Liu:2022ajh}, gluon thrust \cite{Beneke:2022obx}, and $e\mu$ backward scattering \cite{Bell:2022ott}, $B_c\to \eta_c$ form factors \cite{Bell:2024bxg}, etc.
Below we will provide an analytic calculation of the leading logarithms at both NLO and NNLO by using a method developed in Refs. \cite{Wang:2019mym,Hou:2025ovb}.
This would not only serve as a nontrivial check of the above result from differential equations, but also may help to develop a formalism for the resummation of such logarithms to all orders.

\begin{figure}[ht]
	\centering
	\begin{minipage}{0.18\linewidth}
		\centering
		\includegraphics[width=1\linewidth]{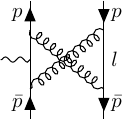}
		\caption*{(a)}
	\end{minipage}
 	\centering
	\begin{minipage}{0.18\linewidth}
		\centering
		\includegraphics[width=1\linewidth]{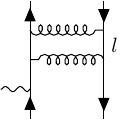}
		\caption*{(b)}
	\end{minipage}
 	\centering
	\begin{minipage}{0.18\linewidth}
		\centering
		\includegraphics[width=1\linewidth]{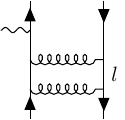}
		\caption*{(c)}
	\end{minipage}
 	\centering
	\begin{minipage}{0.18\linewidth}
		\centering
		\includegraphics[width=1\linewidth]{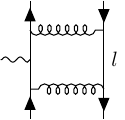}
		\caption*{(d)}
	\end{minipage}
 	\centering
	\begin{minipage}{0.18\linewidth}
		\centering
		\includegraphics[width=1\linewidth]{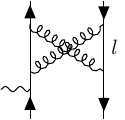}
		\caption*{(e)}
	\end{minipage}
\caption{One-loop diagrams containing double logarithms via a soft quark loop.}
\label{fig:one_loop}
\end{figure}

The large logarithms appear for the first time at one-loop.
The double logarithms can arise when the gluon momentum becomes soft and (anti-)collinear; see Ref. \cite{terHoeve:2023ehm} for the analyses of the massive form factor with the method of regions \cite{Smirnov:1997gx,Beneke:1997zp}.
This contribution is proportional to the color charge of the external collinear particle and thus vanishes after summing over all diagrams, since the mesons are color neutral.
The other kind of region that contributes to the double logarithms involves the soft quark.
The relevant Feynman diagrams are shown in Fig. \ref{fig:one_loop}.

Take diagram \ref{fig:one_loop}(d) as an example.
The loop momentum of the quark is denoted by $l$.
The collinear and anti-collinear directions are labeled by 
$ n^\mu=(1,0,0,1),$ and $\bar{n}^\mu=(1,0,0,-1)$, respectively.
Any momentum $k$ can be represented by $(k^+, k^- , k_{\perp}^{\mu})$ with $k^+=n\cdot k$ and $k^-= \bar{n}\cdot k$.
The external charm (anti-)quarks have collinear $p\sim (r,1,0)\sqrt{s}$ or anti-collinear $\bar{p}\sim (1,r,0)\sqrt{s}$ momenta.
The double logarithms receive contributions from four regions: the hard region $l\sim (1,1,1)\sqrt{s}$, the collinear region $l\sim (r,1,\sqrt{r})\sqrt{s}$, the anti-collinear region $l\sim (1,r,\sqrt{r})\sqrt{s}$ and the soft  region $l\sim (\sqrt{r},\sqrt{r},\sqrt{r})\sqrt{s}$.
The soft region gives the amplitude (with the LO amplitude split)
\begin{align}
\mathcal{A}_s 
%& = i\pi   \alpha_s  C_F\mu^{2\epsilon}\int\frac{d^Dl}{(2\pi)^D} \frac{(n\cdot l)(\bar n\cdot l)-l^2+3m^2}{(l^2-m^2)(n\cdot l)^2(\bar n\cdot l)^2} \nonumber \\
=i\pi   \alpha_s  C_F\mu^{2\epsilon}\int\frac{d^Dl}{(2\pi)^D} \frac{1}{(l^2-m_c^2)l^+  l^-} ,
\end{align}
where we have neglected terms that do not contribute to the double logarithm.
The Feynman prescription for the three denominators is given by $+i\varepsilon,+i\varepsilon,-i\varepsilon$, respectively. 
Naive integration leads to endpoint divergences even with a dimensional regulator.
Following the method in Ref. \cite{Hou:2025ovb},  
we add auxiliary $\delta$-regulators in the full denominators, i.e., $(l+p)^2\to (l+p)^2+\delta_1 s$ and $(l-\bar{p})^2\to (l-\bar{p})^2+\delta_2 s$.
Consequently, the above denominators in the soft region are modified as $l^+\to l^++2\delta_1\sqrt{s}, l^-\to l^--2\delta_2\sqrt{s}$.
Integrating over $l^-$ using the residue theorem, we then obtain
\begin{align}
\mathcal{A}_s & =\frac{\alpha_sC_F\mu^{2\epsilon}      }{16\pi}\int^{\infty}_0 dl_T^2 \nonumber\\
& \int_{0}^{\infty} d l^+ \frac{l_T^{-2\epsilon}}{l_T^2+m_c^2-2\delta_2 \sqrt{s} l^+}\frac{1}{ l^+ +2\delta_1\sqrt{s}}.
\label{eq_1}
\end{align}
The endpoint divergences at $l^+=0$ and $l^+=\infty$ are now regulated by $\delta_1$ and $\delta_2$, respectively.
The integration of $l_T^2$ generates an ultraviolet divergence in $1/\epsilon$.
The collinear and anti-collinear regions have the same structure.
All the ultraviolet divergences would cancel with the infrared divergences in the hard region.
Given this insight, we split the $l_T^2$ integration into two parts, $(0,s)$ and $(s,\infty)$.
In the second part, $l_T$ is considered as a hard scale,
and thus the small scale $m_c$ can be dropped.
As a result, no logarithms would appear in this part.
The first part contains no dimensional divergences and is given by
\begin{align}
\mathcal{A}_{s}^{l_T^2\le s}
=\frac{\alpha_sC_F       }{32\pi}\text{ln}(r)\left[  2 \text{ln}(\delta_1) +2\text{ln}(\delta_2) -\text{ln}(r)             \right] \,,
\end{align}
where we have kept only the double logarithms.
Similarly, we obtain the results in the collinear and anti-collinear regions,
\begin{align}
\mathcal{A}_{ c}^{ l_T^2 \le s}=&\frac{ \alpha_sC_F       }{16\pi}\text{ln}(r)\text{ln}(\delta_2)\,,
\\\mathcal{A}_{\bar c}^{ l_T^2 \le s}=&\frac{ \alpha_sC_F     }{16\pi}\text{ln}(r)\text{ln}(\delta_1)\,.
\end{align}
Note that the expansion of $\mathcal{A}_{c}$ in the soft (or anti-collinear) region reproduces $\mathcal{A}_{s}$.
Therefore, a subtraction needs to be performed to avoid double-counting \cite{Manohar:2006nz}.
Taking the sum of the above contributions yields
\begin{align}
&(\mathcal{A}_c^{l_T^2\le s}-\mathcal{A}_s^{l_T^2\le s})+(\mathcal{A}_{\bar c}^{l_T^2\le s}-\mathcal{A}_s^{l_T^2\le s})+\mathcal{A}_s^{l_T^2\le s}
\nonumber\\
=&\frac{\alpha_sC_F}{32\pi}  \text{ln}^2(r) \,,
\end{align}
which agrees with the expansion of the full one-loop amplitude.
It can be seen that all the $\delta$-regulators cancel out.
Another important feature of this method is that the collinear and anti-collinear results are proportional to $\ln(\delta_i)$ and thus can be taken to be vanishing if setting $\delta_i=1$.
This indicates a shortcut to obtain the leading logarithms.
One needs to calculate only the soft region contribution analytically.

\begin{figure}[ht]
	\centering
	\begin{minipage}{0.22\linewidth}
		\centering
		\includegraphics[width=1\linewidth]{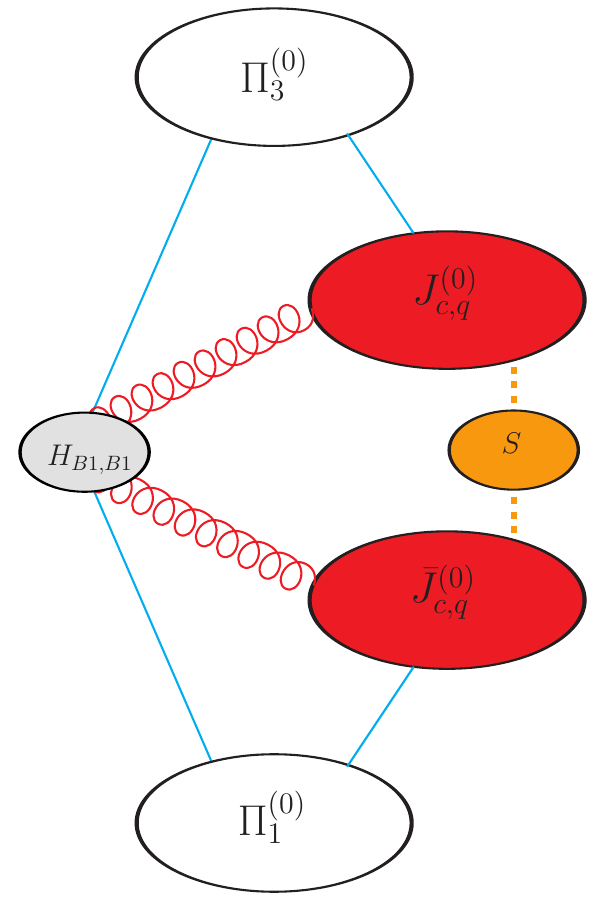}
		\caption*{(a)}
	\end{minipage}
 	\centering
	\begin{minipage}{0.22\linewidth}
		\centering
		\includegraphics[width=1\linewidth]{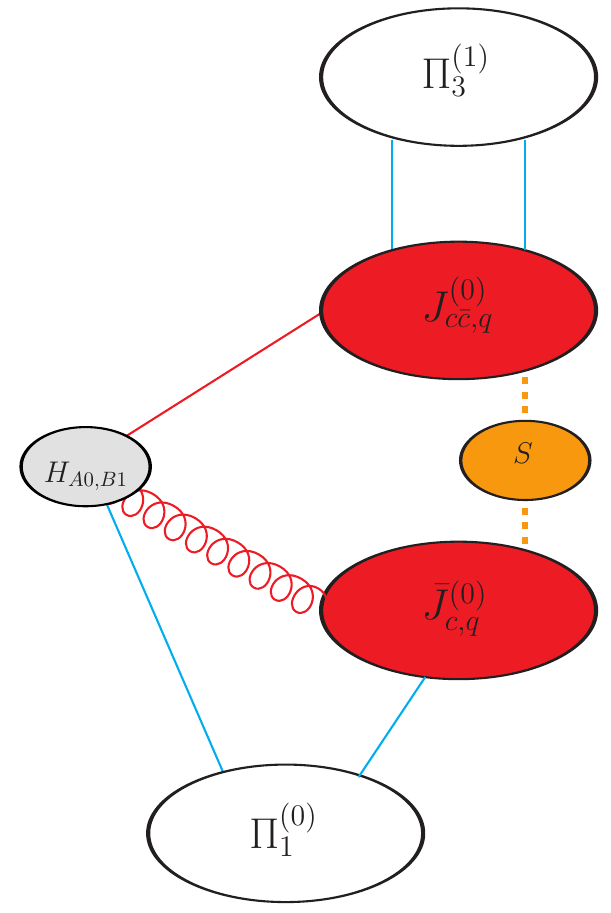}
		\caption*{(b)}
	\end{minipage}
 	\centering
	\begin{minipage}{0.22\linewidth}
		\centering
		\includegraphics[width=1\linewidth]{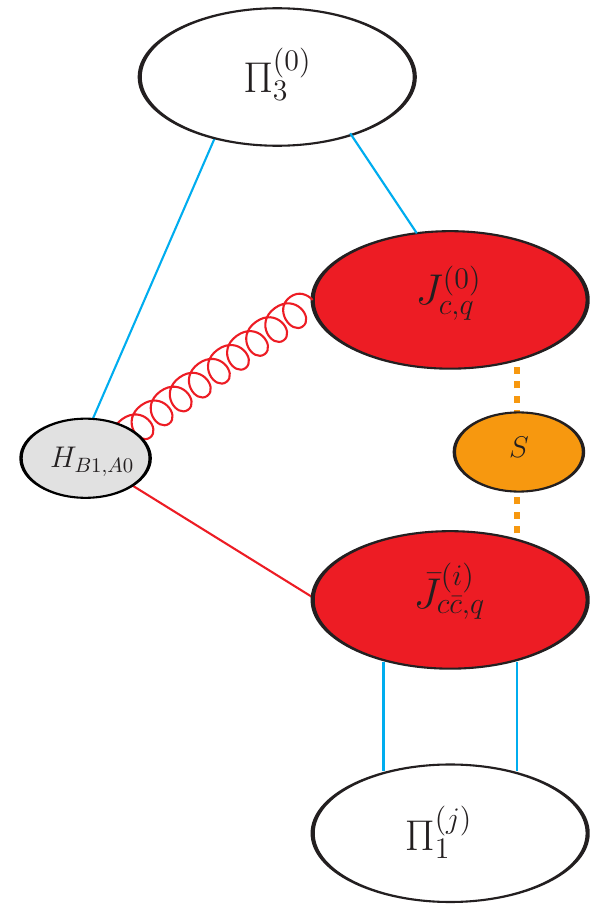}
		\caption*{(c)}
	\end{minipage}
 	\centering
	\begin{minipage}{0.22\linewidth}
		\centering
		\includegraphics[width=1\linewidth]{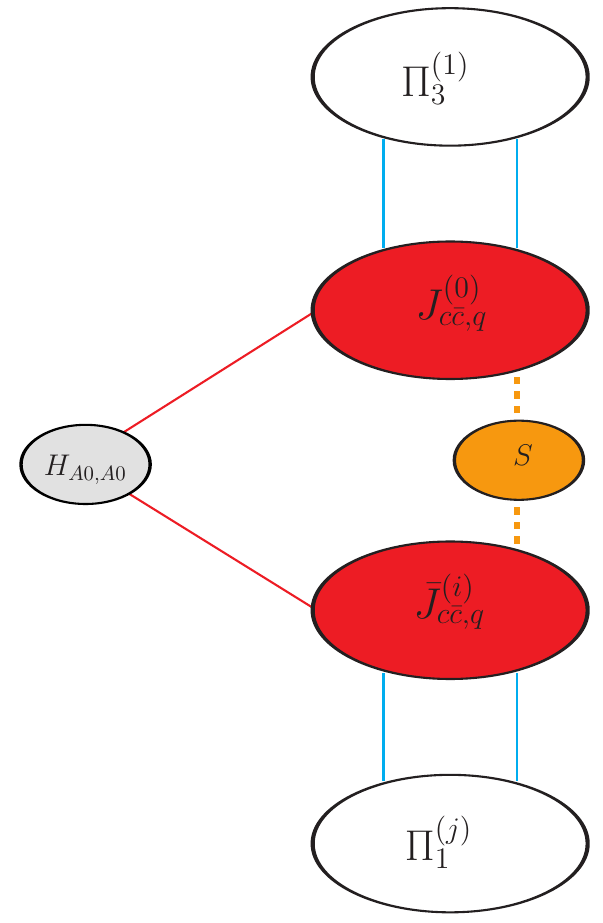}
		\caption*{(d)}
	\end{minipage}
\caption{Reduced diagrams contributing to double logarithms via a soft quark loop. The superscripts satisfy $i+j=1$. The  soft and collinear Wilson lines are not shown explicitly.}
\label{fig:one_loop_topo}
\end{figure}

The other diagrams in Fig. \ref{fig:one_loop} can be computed similarly.
Based on these analyses of regions, we show the reduced diagrams contributing to double logarithms in Fig. \ref{fig:one_loop_topo}.
The whole perturbative amplitude can be considered as a convolution of the hard function, the radiative jet function, and the soft function, which is interfaced with projectors.
The hard function represents the contribution of all hard propagators with $\mathcal{O}(1)$ off-shellness.  It can be considered as the Wilson coefficients of the effective operators, consisting of the building blocks $W_{hc}^{\dagger}\xi_{hc}$ and $W_{hc}^{\dagger}[iD_{\perp}^{\mu}W_{hc}]$ \cite{Beneke:2017ztn,Beneke:2018rbh}.
The radiative jet function $J_{c,q}$ describes the transition of a hard collinear gluon to a collinear anti-charm-quark and a soft charm-quark, denoted by its subscripts $c$ and $q$, respectively.
The other radiative jet function $J_{c\bar{c},q}$ describes the transition of a hard collinear charm-quark to a collinear (anti-)charm-quark pair and a soft charm-quark.
The soft function $S$ incorporates the effects of all soft particles, including the soft Wilson lines which are not shown explicitly in the figures.
The external collinear (anti-)charm-quark pair forms a state of spin-singlet (triplet) via the projectors \cite{Bodwin:2014dqa}
\begin{subequations}
\begin{align}
&\Pi_1(\bar p, \bar p)=\Pi_1^{(0)}+\Pi_1^{(1)}+\mathcal{O} (r^{1/2})\,,
\\& \Pi_3(p, p,\lambda)=\Pi_3^{(0)}+\Pi_3^{(1)}+\mathcal{O} (r^{1/2})\,,
\end{align}
\end{subequations}
where we have performed the power expansion in $r$ and the first two terms are given by
\begin{subequations}
\begin{align}
&\Pi_1^{(0)}=-\frac{\sqrt{s}}{8\sqrt{2}m_c}\gamma_5\slashed{\bar n}\,, \ \ \ \ \Pi_1^{(1)}=-\frac{1}{2\sqrt{2}}\gamma_5\,,
\\&\Pi_3^{(0)}=-\frac{\sqrt{s}}{8\sqrt{2}m_c}\slashed{\epsilon}^{*}_{\perp}(\lambda)\slashed{n}\,, \  \Pi_3^{(1)}=-\frac{1}{2\sqrt{2}}\slashed{\epsilon}^{*}_{\perp}(\lambda)\,.
\end{align}
\end{subequations}

In Fig. \ref{fig:one_loop_topo}(a),  the radiative jet and soft functions are of order $r^{-1/2}$ and $r^{1/2}$, respectively,
and thus the soft loop momentum convolution gives a result of order $r^{1/2}$.
The two leading power projectors are of order $r^{-1}$.
So the amplitude of this diagram is of order $r^{-1/2}$, same as the LO amplitude.
It is also interesting to note that the Dirac structures of $\Pi_1^{(0)}$ and $\Pi_3^{(0)}$ require a helicity flip along the fermion line,
which can only occur in the soft function since the radiative jet functions are at leading power.
In Fig. \ref{fig:one_loop_topo}(b), the projector $\Pi_3^{(1)}$ is power suppressed.
However, $J_{c\bar{c},q}^{(0)}$ is power enhanced by $r^{-1/2}$ compared to $J_{c,q}^{(0)}$.
Thus, the amplitude is at the same power as Fig. \ref{fig:one_loop_topo}(a).
Fig. \ref{fig:one_loop_topo}(c) contains two types of contributions.
The first type, with $i=0$ and $j=1$, is analogous to Fig. \ref{fig:one_loop_topo}(b) with the collinear and anti-collinear directions exchanged.
The other type,  with $i=1$ and $j=0$, is similar to Fig. \ref{fig:one_loop_topo}(c), with the replacement of $\bar{J}_{c,q}^{(0)}$ by $\bar{J}_{c\bar{c},q}^{(1)}$, which is of the same power.
The power counting of Fig. \ref{fig:one_loop_topo}(d) can be understood following the same logic.

\begin{figure}[ht]
	\centering
	\begin{minipage}{0.33\linewidth}
		\centering
		\includegraphics[width=1\linewidth]{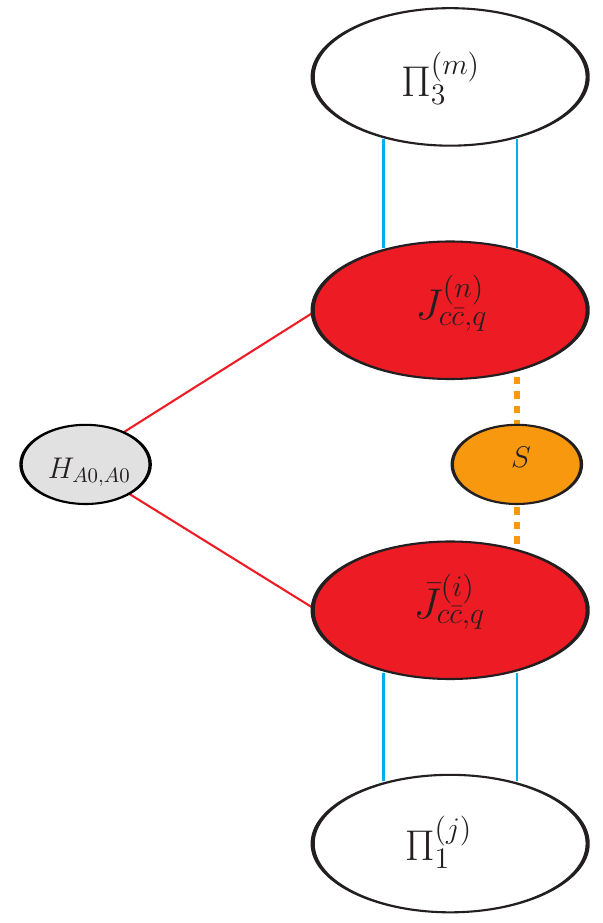}
		\caption*{(a)}
	\end{minipage}
 	\centering
	\begin{minipage}{0.33\linewidth}
		\centering
		\includegraphics[width=1\linewidth]{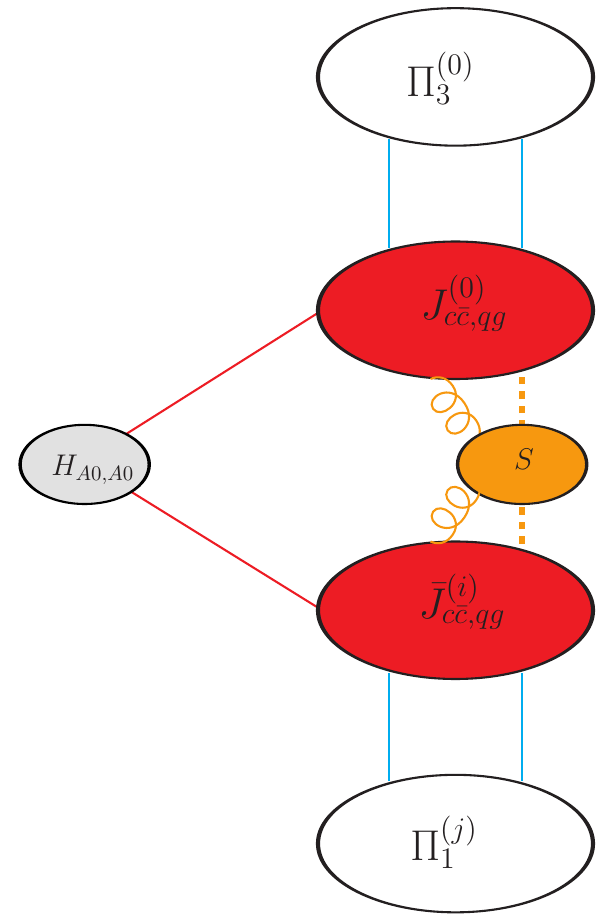}
		\caption*{(b)}
	\end{minipage}
 	\centering
\caption{Sample reduced diagrams contributing to leading logarithms at the two-loop level. The superscripts satisfy $i+j=1$ and $m+n=1$. The soft and collinear Wilson lines are not shown explicitly.}
\label{fig:two_loop_topo}
\end{figure}

At the two-loop level, there are additional types of contributions.
Sample reduced diagrams are shown in Fig. \ref{fig:two_loop_topo}.
Fig. \ref{fig:two_loop_topo}(a) generalizes Fig. \ref{fig:one_loop_topo}(d) to include the case with leading power projector $\Pi_3^{(0)}$ but with subleading radiative jet function $J_{c\bar{c},q}^{(1)}$.
Fig. \ref{fig:two_loop_topo}(b) introduces a new type of radiative jet function, $J_{c\bar{c},qg}$, that emits both a soft quark  and a soft physical gluon.
It can be considered as obtained by inserting in $J_{c\bar{c},q}$ a subleading power interaction ($\mathcal{O}(r^{1/4})$) involving a soft gluon.
The suppression induced by $J_{c\bar{c},qg}$ is balanced by the enhancement of the leading power projector $\Pi_3^{(0)}$, compared to Fig. \ref{fig:one_loop_topo}(d).
The new factorization topologies appearing at higher loops represent a challenge to the development of a resummation formalism in this process.

We calculated most of the two-loop diagrams contributing to the leading logarithms with the method described above.
Although the soft quark momentum mode is fixed, the additional gluon loop momentum can have hard, (anti-)hard-collinear, (anti-)collinear, (anti-)soft-collinear (scaling as $(r^{3/2},r^{1/2},r)$), and soft modes.
The (anti-)soft-collinear mode does not appear in the reduced diagrams because its contribution cancels out in the amplitude due to the same reason as the soft gluon at one loop.
We performed the calculation loop-by-loop.
The additional loop was calculated analytically using the Feynman parameterization.
Thanks to the $\delta$-regulators and the cut on $l_T^2$ applied to the remaining soft quark loop, the result can be safely expanded in $\epsilon$.
Then the integration over $l$ was carried out using the residue theorem.
%\cqcom{[How many are needed in total?]}
We investigated 80 Feynman diagrams,  finding agreement with the direct computation.
It is crucial to compare the subset of diagrams in which the leading logarithms induced by pure soft gluons cancel.

However, there exist a few diagrams that cannot be computed using the $\delta$-regulators.
These diagrams contain the tent-type subdiagram, which has also appeared in the muon-electron backward scattering \cite{Bell:2022ott}.
The structure of the involved two-dimensional endpoint divergences requires advanced applications of the  $\delta$-regulators (or other regulators).
This interesting problem is left to future work.
Nevertheless, the region analyses expose the intricate factorization structure hidden behind the leading logarithms and may shed light on the resummation of these logarithms to all orders.

\section{Phenomenology}
To facilitate comparison between theoretical predictions and experimental results, we summarize the input parameters used in this work in Tab.~\ref{table:nlf}. 
The masses of charm quark and bottom quark are taken as $m_c=1.5$ GeV and $m_b=4.8$ GeV, respectively, while other light quarks are considered as massless in this paper. $N_{lf}$ denotes the number of active flavors of quarks, while $N_{lp}$ and $N_{lm}$ denote the number of active positive-charged and negative-charged quarks, respectively. $|R_{H}(0)|^2$ is the long-distance matrix element for the quarkonium $H$ in the NRQCD framework, which was taken from Refs. \cite{Zhang:2005cha,Eichten:1995ch,Brambilla:2022ayc}. At the lowest order in velocity expansion, we have $|R_{J/\psi}(0)|^2=|R_{\eta_c}(0)|^2$ and $|R_{\Upsilon}(0)|^2=|R_{\eta_b}(0)|^2$.
%We take the relation between LDME and  wave function as $|R_{H}(0)|^2=\frac{\mathrm{N}_c}{2\pi}\langle\mathcal{O}\rangle_{H}$.

\begin{table}[h!]
\renewcommand\arraystretch{2}
\setlength{\belowcaptionskip}{0.2cm}
\centering
{
  \begin{tabular}{cccccc}
    \hline
& $m_{Q}/\mathrm{GeV}$      & $N_{lf}$   & $N_{lp}$       & $N_{lm}$      &$|R_{H}(0)|^2/\mathrm{GeV}^3$ \\
    \hline
$ J/\psi +\eta_c $          & 1.5     &3         &1          &2           &0.978 \\
    \hline
$\Upsilon +\eta_b $         & 4.8     &4         &2           &2          &6.477     \\
    \hline
\end{tabular}
}
\caption{Different  input values taken for charm/bottom quarkonium production processes.}\label{table:nlf}
\end{table}

The QCD coupling constant $\alpha_s(\mu_R) $  is evaluated using our in-house program with $m_c,m_b$ and $\mu_R$ as input, yielding, for instance, $\alpha_s(91.2~\text{GeV})=0.118$ and $\alpha_s(10.58~\text{GeV}) = 0.177$.  Throughout this paper, the factorization scale $\mu_\Lambda$ is set to be $m_Q$. The QED coupling constant is fixed at $\alpha=1/137$.  We also vary the quark mass by $\pm 0.2~\text{GeV}$ to analyze its uncertainty.

\subsection{$J/\psi+\eta_c$ production}
%\begin{table*}[htb]
%\renewcommand\arraystretch{2}
%\centering
%\begin{tabular}{|c|c|c|c|}
%\hline
%\textbf{$\sqrt{s}$ (GeV)} & \textbf{LO (fb)} & \textbf{NLO (fb)} & \textbf{NNLO (fb)} \\
%\hline
%10.52 &
%\quad$5.62^{+2.41+1.04}_{-1.70-1.12}$ \quad &
%\quad$11.88^{+3.79+3.27}_{-3.02-2.96}$ \quad &
%\quad$17.53^{+4.07+6.96}_{-3.69-5.50}$ \quad \\
%\hline
%10.58 &
%\quad$5.40^{+2.34+0.98}_{-1.63-1.05}$ \quad &
%\quad$11.43^{+3.69+3.10}_{-2.90-2.82}$ \quad &
%\quad$16.90^{+3.99+6.64}_{-3.57-5.25}$ \quad \\
%\hline
%\end{tabular}
%\caption{The cross sections for the process $e^{+} e^{-} \rightarrow J/\psi + \eta_c$ at different energies $\sqrt{s}$. The central values are evaluated at $\mu_R = \sqrt{s}/2$ and $m_c = 1.5$~GeV. The first uncertainty arises from varying the renormalization scale $\mu_R$ from $3~\text{GeV}$ to $\sqrt{s}$, and the second from varying $m_c$ from $1.3~\text{GeV}$ to $1.7~\text{GeV}$. The factorization scale $\mu_\Lambda=m_c=1.5~\text{GeV}$.}\label{tab:JE-CS-0}
%\end{table*}
We present in Tab.~\ref{tab:JE-CS-1} the total cross sections at various collider energies. The cross section at $\sqrt{s}=10.58$~GeV changes from $13.87$ fb to $11.85$ fb when varying $\mu_R$ from $\sqrt{s}/2$ to $\sqrt{s}$,  consistent with experimental measurements in Eqs.~\eqref{belle} and \eqref{babar}.
The cross section at $\sqrt{s}=10.52$~GeV is $14.36$ fb and $12.29$ fb when setting $\mu_R$ at $\sqrt{s}/2$ and $\sqrt{s}$, respectively, serving as a prediction for the upcoming Belle II measurement. Finally, the total cross sections at $\sqrt{s}=91.2$~GeV and 250~GeV are respectively $8.48 \times 10^{-7}$~fb and $2.84\times 10^{-10}$~fb, which could be of interest for future colliders.

\begin{figure}[htb]
\centering{
  \includegraphics[width=0.9\linewidth]{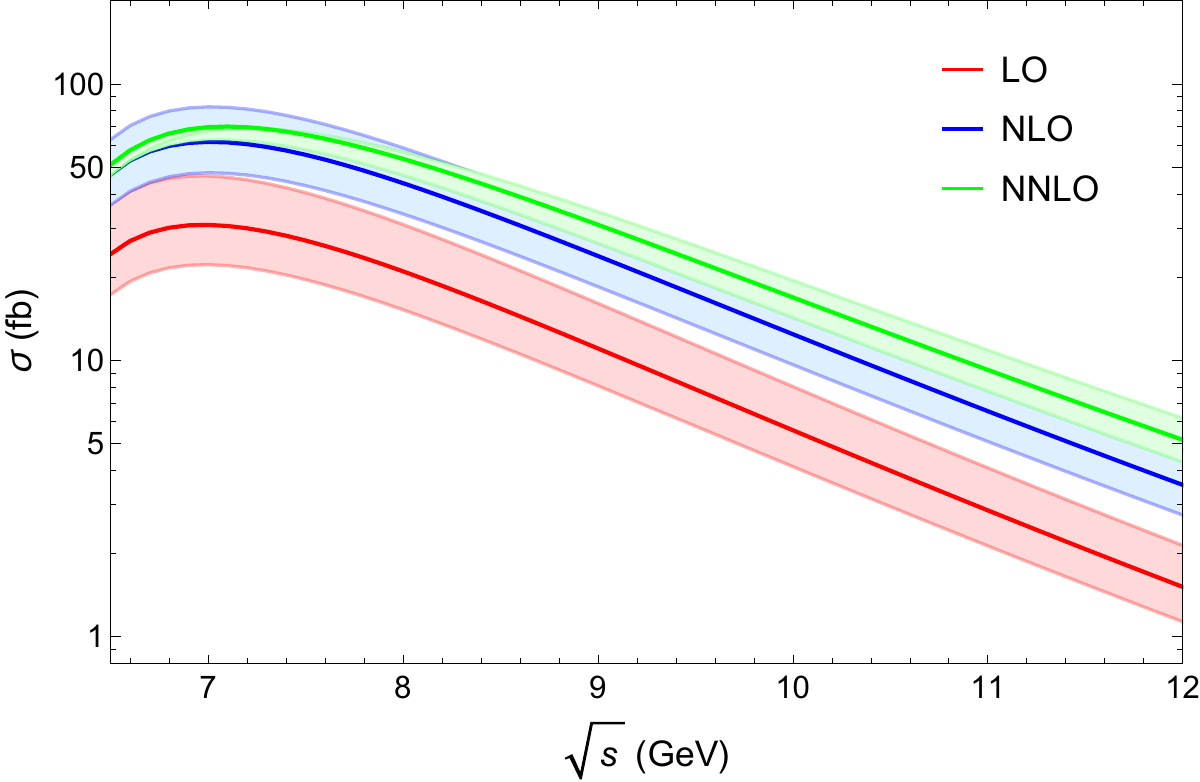}
  \caption{The LO, NLO, NNLO total cross sections for $e^+e^-\rightarrow \gamma^{\ast} \rightarrow J/\psi +\eta_c$ with the colliding energy ranging from 6.5~GeV to 12~GeV. The red, blue, and green bands stand for LO, NLO, and NNLO cross sections. The upper edge, central solid line, and lower edge correspond to $\mu_R=\sqrt{s}/2, \sqrt{s}, 2\sqrt{s}$, respectively. }\label{fig:csJErange1}
  }
\end{figure}

Fig.~\ref{fig:csJErange1} shows the total cross section for $J/\psi + \eta_c$ production, at the range of $\sqrt{s}=[6.5,12]$~GeV. The red, blue, and green bands stand for the LO, NLO, and NNLO cross sections, respectively. The width of each band represents the theoretical uncertainty from the renormalization scale variation, evaluated between $\mu_R = \sqrt{s}/2$ (upper edge) and $\mu_R = 2\sqrt{s}$ (lower edge). A logarithmic scale is used for the vertical axis to clearly display the behavior near the physically interesting region around 10.6 GeV, given the cross section's steep decline. The cross section peaks near 7 GeV, a value slightly above four times $m_c$. This comparison among the LO, NLO and NNLO results clearly demonstrates the significant contribution of the $\mathcal{O}(\alpha_s^2)$ correction.
%These curves are threefold, where the width of the band stands for the renormalization scale  $\mu_R$ dependence, wherein the upper edge is for $\sqrt{s}/2$ and the bottom edge for  $2\sqrt{s}$. Due to the steep decline of the cross section, a logarithmic scale is used to highlight the region around 10.6~GeV. The cross section peaks at about 7~GeV, which is slightly above four times the charm quark pole mass. The simple comparison manifests the importance of the $\mathcal{O}(\alpha_s^2)$ correction.
%The large uncertainties due to different choices of $\mu_R$ indicate that even higher order perturbative corrections are not negligible.

\begin{figure}[htb]
\centering{
  \includegraphics[width=0.9\linewidth]{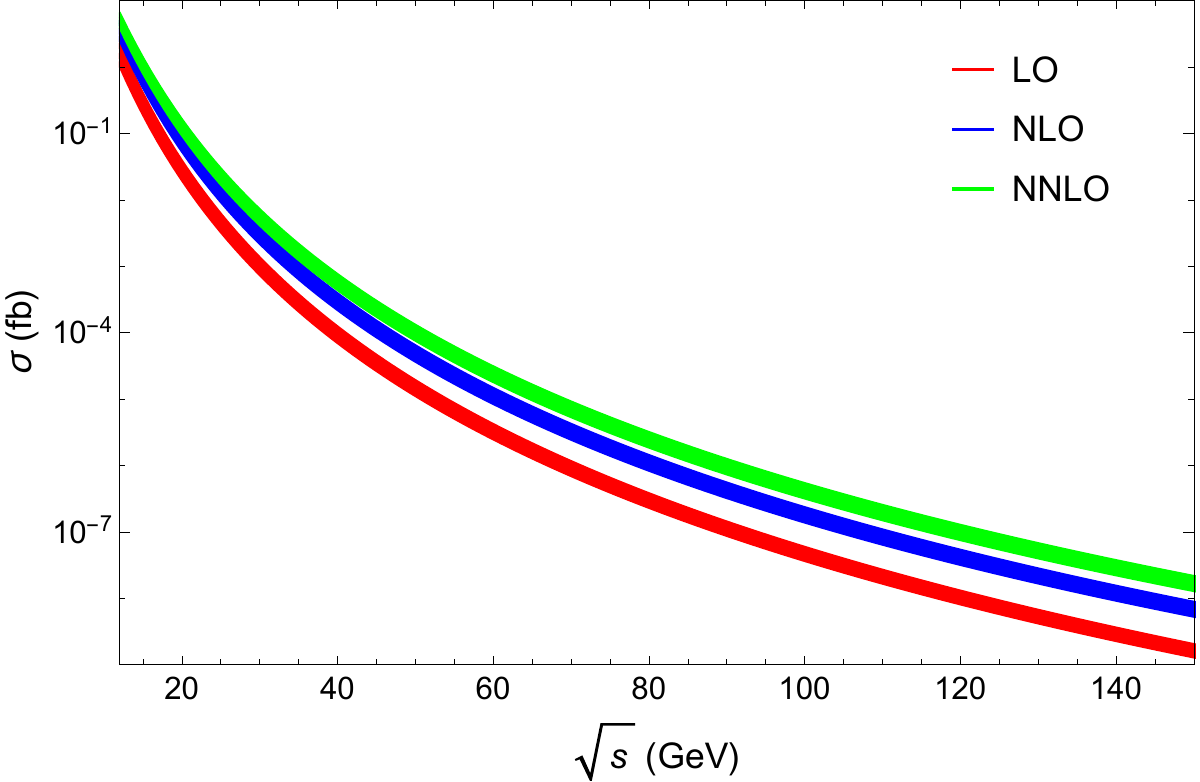}
  \caption{Same as Fig. \ref{fig:csJErange1} but with the colliding energy ranging from 12~GeV to 150~GeV. }\label{fig:csJErange2}
  }
\end{figure}

The total cross section over a broader energy range is presented on a logarithmic scale in Fig.~\ref{fig:csJErange2}. The cross section falls drastically with increasing energy. In this high-energy regime, higher-order corrections become increasingly significant due to the presence of large logarithmic terms.

%Fig.~\ref{fig:csJErange2} presents the total cross section for $J/\psi + \eta_c$ production over a broader range, displayed on a logarithmic scale. It is evident that the cross section drops drastically at high energy region. And at the same time, higher order corrections become even significant at high energy region, due to the logarithmic terms. 
%, significantly faster than $1/s^4,$ due to the large log terms.  

\begin{table*}[htb]
\renewcommand\arraystretch{2}
\centering
\begin{tabular}{|c|c|c|c|}
\hline
\textbf{$\sqrt{s}$ (GeV)} & \textbf{LO (fb)} & \textbf{NLO (fb)} & \textbf{NNLO (fb)} \\
\hline
10.52 & $3.93^{+1.70+0.73}_{-1.02-0.78}$ & $8.86^{+3.02+2.26}_{-1.98-2.11}$ & $12.29^{+2.07+4.50}_{-1.96-3.66}$ \\
\hline
10.58 & $3.77^{+1.63+0.69}_{-0.98-0.74}$ & $8.53^{+2.90+2.15}_{-1.91-2.01}$ & $11.85^{+2.02+4.29}_{-1.90-3.49}$ \\
\hline
22.5 
& $\left(1.05^{+0.36+0.03}_{-0.23-0.03}\right) \times 10^{-2}$ 
& $\left(2.87^{+0.92+0.24}_{-0.61-0.22}\right) \times 10^{-2}$ 
& $\left(5.01^{+1.34+0.75}_{-0.96-0.62}\right) \times 10^{-2}$ 
\\
\hline
91.2 
& $\left(9.83^{+2.46+0.02}_{-1.78-0.02}\right) \times 10^{-8}$ 
& $\left(3.73^{+1.04+0.19}_{-0.73-0.16}\right) \times 10^{-7}$ 
& $\left(8.48^{+2.34+0.75}_{-1.66-0.64}\right) \times 10^{-7}$ 
\\
\hline
250 
& $\left(2.33^{+0.50+0.00}_{-0.37-0.00}\right) \times 10^{-11}$ 
& $\left(10.87^{+2.73+0.47}_{-1.98-0.40}\right) \times 10^{-11}$ 
& $\left(2.84^{+0.74+0.22}_{-0.53-0.18}\right) \times 10^{-10}$ 
\\
\hline
\end{tabular}
\caption{The cross sections for $e^+e^- \to J/\psi + \eta_c$ at different colliding energies. The central values are evaluated at $\mu_R = \sqrt{s}$ and $m_c = 1.5$~GeV. The first uncertainty is estimated by varying $\mu_R$ from $\sqrt{s}/2$ to $2\sqrt{s}$, and the second estimated by varying $m_c$ from $1.3$~GeV to $1.7$~GeV.The factorization scale $\mu_\Lambda=m_c=1.5~\text{GeV}$.}\label{tab:JE-CS-1}
\end{table*}

%Numerical result for cross section，where the renormalization scale is set as $\mu=\sqrt{s}/2$ and factorization scale $\mu_{\Lambda}=m_c=1.5~GeV$

\subsection{$\Upsilon+\eta_b$ production}
Our analytical results and numerical asymptotic expansion retain explicit dependence on heavy quark mass, which makes them readily applicable to the process for the bottomonium counterpart.  By substituting the relevant bottom quark parameters, as listed in Tab.~\ref{table:nlf}, we obtain predictions for the cross section of $\Upsilon+\eta_{b}$ production. 

From the numerical results shown in Tab.~\ref{tab:UE-CS-1}, we observe that the absolute magnitude of the cross section at $\sqrt{s}=22.5$~GeV is comparable to that of $J/\psi+\eta_{c}$.
It can be seen from Tab.~\ref{tab:JE-CS-1} and \ref{tab:UE-CS-1}, the relative error of $J/\psi+\eta_{c}$ are generally larger than that of $\Upsilon+\eta_{b}$. For example, at $\sqrt{s}=$91.2~GeV, the relative uncertainty is approximately  28\% for  $J/\psi+\eta_{c}$ and 15\% for $\Upsilon+\eta_{b}$.  Additionally, for both processes, the relative error from the renormalization scale and the quark pole mass tend to decrease at higher energies.

\begin{figure}[htb]
\centering{
\includegraphics[width=0.9\linewidth]{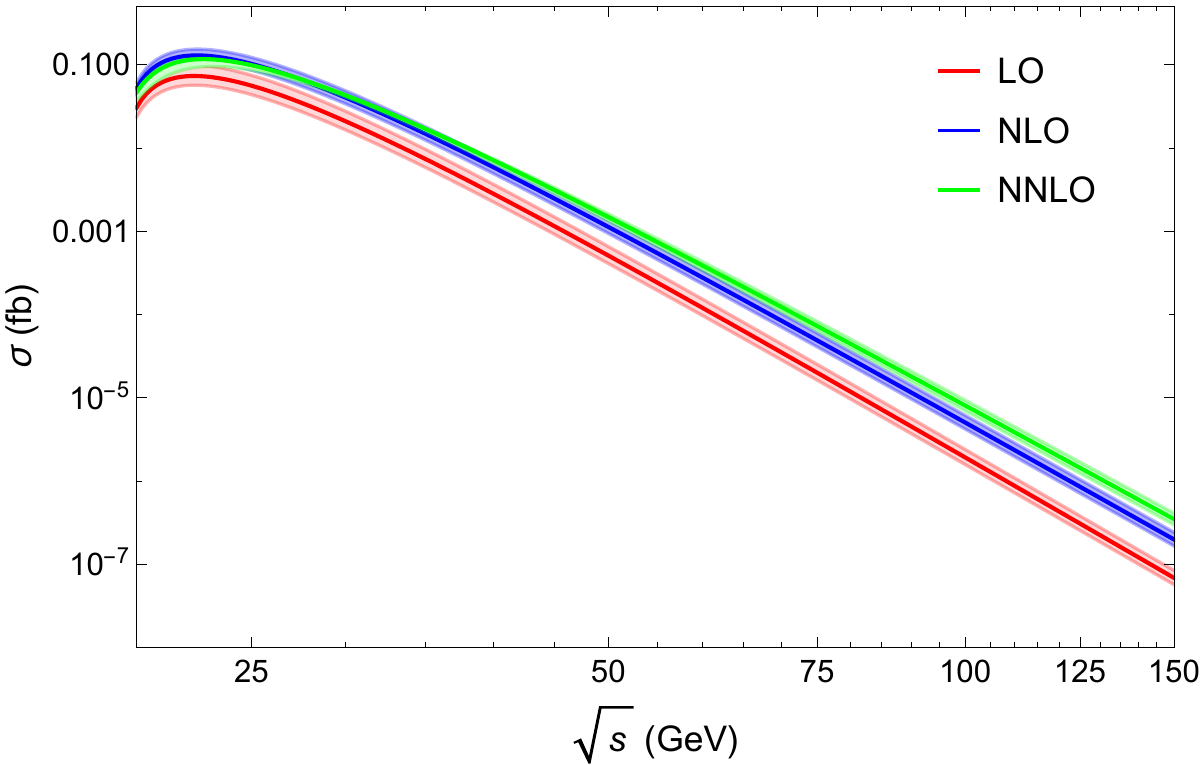}
\caption{The LO, NLO, NNLO total cross sections for $e^+e^-\rightarrow \gamma^{\ast} \rightarrow \Upsilon +\eta_b$ when $\sqrt{s}$ varies from 20~GeV to 150~GeV. The red, blue, and green bands stand for LO, NLO, and NNLO cross sections. The upper edge, central solid line, and lower edge correspond to $\mu_R=\sqrt{s}/2, \sqrt{s}, 2\sqrt{s}$, respectively.}\label{fig:csUErange1}
  }
\end{figure}

Fig.~\ref{fig:csUErange1} shows the cross section of $\Upsilon+\eta_{b}$ in the energy region from 20 to 150~GeV. Both axes are plotted on a logarithmic scale to enhance visibility of the high-energy region. Each band represents the variation associated with the renormalization scale $\mu_R$, where the upper and lower edges correspond to $\mu_R = \sqrt{s}/2$ and $2\sqrt{s}$, respectively.
The cross section peaks around 22.5~GeV, where the  cross section is
\begin{equation}
 \sigma=0.115^{-0.023+0.048}_{+0.002-0.041}~\text{fb}. 
 %\sigma=(4.68_{-0.249-1.66}^{-0.255+1.93}) \times 10^{-2} ~\text{fb}. 
\end{equation}
As shown in Fig.~\ref{fig:csUErange1},  when $\sqrt{s}<30$ GeV, the NNLO corrections (the $\mathcal{O}(\alpha_s^2)$ terms) contribute a relatively small fraction to the total cross section; this is visually manifested by the NNLO band (in green) being almost entirely covered by the NLO band (in blue). This reflects significantly  better perturbative convergence compared to the $J/\psi +\eta_c$ case. Moreover, the uncertainty originating from the renormalization scale  $\mu_R$ is noticeably reduced compared to NLO. However, when  $\sqrt{s}>30$ GeV, NNLO corrections become more significant, which should be caused by large logarithms. 
It is worth noting that in the region where $\sqrt{s} \gg 4m_Q$, the cross section of $\Upsilon+\eta_{b}$ is much larger than that of $J/\psi +\eta_c$ due to the relatively larger long-distance matrix element.

%\begin{figure}[htb]
%\centering{
%\includegraphics[width=0.9\linewidth]{figures/UEfig2.pdf}
%\caption{The LO, NLO and NNLO cross section for $e^+e^-\rightarrow \gamma^{\ast} \rightarrow \Upsilon +\eta_b$ ranging from 40~GeV to 150~GeV. The bands are threefold, from lower edge to upper edge, representing $\mu_R=2\sqrt{s},\sqrt{s},\sqrt{s}/2$,  respectively.}\label{fig:csUErange2}}
%\end{figure}

\begin{table*}[htb]
\renewcommand\arraystretch{2}
\centering
\begin{tabular}{|c|c|c|c|}
\hline
\textbf{$\sqrt{s}$ (GeV)} & \textbf{LO (fb)} & \textbf{NLO (fb)} & \textbf{NNLO (fb)} \\
\hline
22.5 
& $0.073^{+0.025+0.025}_{-0.016-0.023} $ 
& $0.128^{+0.025+0.047}_{-0.020-0.043} $ 
& $0.115^{-0.023+0.048}_{+0.002-0.041} $ 
\\
\hline
91.2 
& $\left(4.06^{+1.01+0.02}_{-0.73-0.02}\right) \times 10^{-6}$ 
& $\left(10.40^{+2.30+0.21}_{-1.72-0.20}\right) \times 10^{-6}$ 
& $\left(16.28^{+2.51+0.59}_{-2.13-0.56}\right) \times 10^{-6}$ 
\\
\hline
250 
& $\left(1.02^{+0.21+0.00}_{-0.16-0.00}\right) \times 10^{-9}$ 
& $\left(3.31^{+0.72+0.05}_{-0.54-0.05}\right) \times 10^{-9}$ 
& $\left(6.41^{+1.22+0.17}_{-0.96-0.16}\right) \times 10^{-9}$ 
\\
\hline
\end{tabular}
\caption{Cross sections for $e^+e^- \to \Upsilon + \eta_b$ at different $\sqrt{s}$. Central values are evaluated at $\mu_R = \sqrt{s}$ and $m_b = 4.8$~GeV. The first uncertainty is from varying $\mu_R$ from $\sqrt{s}/2$ to $2\sqrt{s}$, and the second from varying $m_b$ from $4.6$~GeV to $5.0$~GeV.}
\label{tab:UE-CS-1}
\end{table*}

\subsection{ Convergence of the expansion in $r$}
%%%%%%%%%%%%%%%%%%
\begin{figure}[!h]
    \begin{minipage}[b]{1.0\linewidth}
        \centering
        \includegraphics[width=1\linewidth]{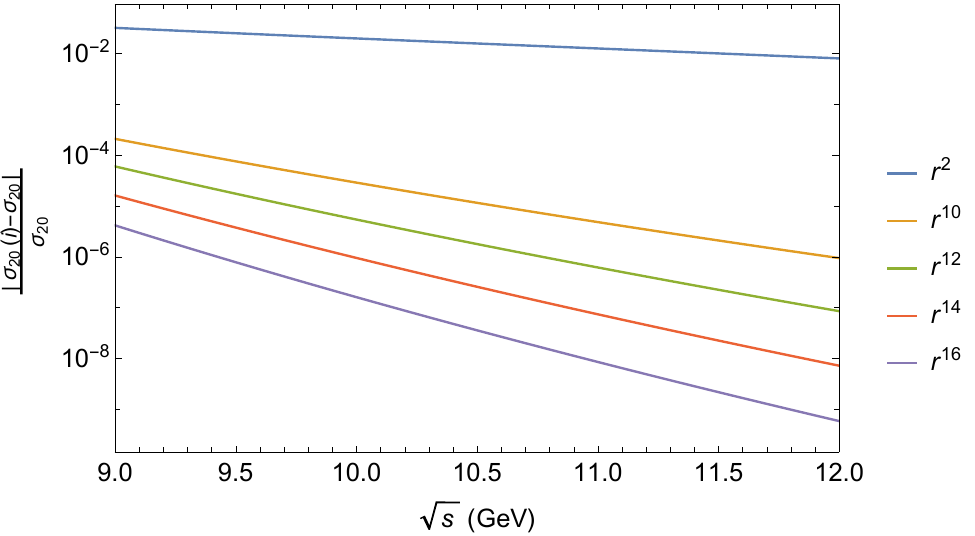}
    \subcaption{%The partial cross section that is propotional to the two-loop amplitudes. 
    The precision of $\sigma_{20}$.}\label{fig:rConverg10}
    \end{minipage}
    \begin{minipage}[b]{0.9\linewidth}
        \centering
        \includegraphics[width=1\linewidth]{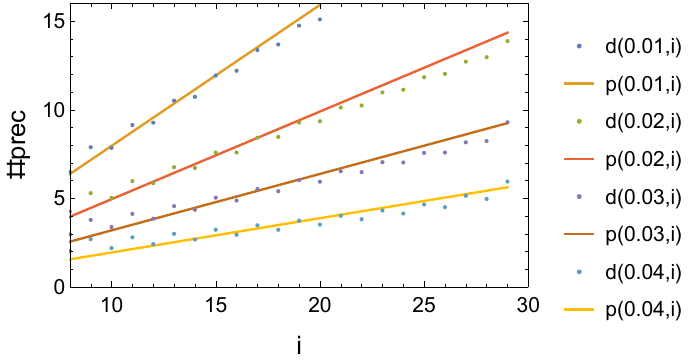}
    \subcaption{The number of precision digits.}\label{fig:rConverg3}
    \end{minipage}
    \caption{Convergence behavior of the two-loop amplitude as an expansion series in $r$.}\label{fig:rConverg}
\end{figure}
%%%%%%%%%%%%%%%%%%
In this subsection, we analyze the convergence behavior of the two-loop amplitude $\mathcal{A}^{\mu(2)}$ within the expansion variable $r$ across different energy regimes. This is done by examining the contribution of the interference term
\begin{equation}
\sigma_{20} \propto \operatorname{Re}\left[\mathcal{A}^{\mu(2)} \mathcal{A}_{\mu}^{(0)*}\right].
\end{equation}
For most phenomenological evaluations in this work, the asymptotic expansion is truncated at $\mathcal{O}(r^{35})$. 

%Fig.~\ref{fig:rConverg1} displays partial contributions to $\operatorname{Re}\left[\mathcal{A}^{\mu(2)} \mathcal{A}^{\mu(0)*}\right]$ with the expansion cut at $r$ order 2,3,8,9 and 10. The curves for $\mathcal{O}(r^{8})$, $\mathcal{O}(r^{9})$, and $\mathcal{O}(r^{10})$ are nearly indistinguishable, demonstrating that truncation at $\mathcal{O}(r^{10})$ already provides excellent convergence and high-precision predictions within this energy range.
Fig.~\ref{fig:rConverg10} displays the precision of partial cross section $\sigma_{20}$ with the asymptotic expansion up to $r^i$ with $i=2,10,12,14,16$. It can be seen that keeping two more powers of expansion brings one more efficient digit of precision. 
To quantify convergence, we define two functions describing the behavior as the expansion order $i$ increases:
\begin{equation}
\begin{aligned}
d(r,i) &= \left| \log_{10} \left| \frac{\sigma_{20}(i) - \sigma_{20}}{\sigma_{20}} \right| \right|, \\
p(r,i) &= \left| \log_{10} \left( \frac{r}{1/16} \right)^i \right|.
\end{aligned}
\end{equation}
Here, $\sigma_{20}(i)$ is the partial cross section evaluated with the two-loop amplitude truncated at $\mathcal{O}(r^{i})$. The function $d(r,i)$ measures the actual precision achieved by the asymptotic expansion. The function $p(r,i)$ represents the predicted precision based on the convergence radius $1/16 = 0.0625$ derived from differential equation theory; it exhibits a linear dependence on $i$.

We analyze the asymptotic expansion for $r = 0.01, 0.02, 0.03$ and $ 0.04$, corresponding approximately to center-of-mass energies of $15.0, 10.61, 8.66$ and $ 7.5$ GeV for $J/\psi + \eta_c$ production. Fig.~\ref{fig:rConverg3} shows excellent agreement between $d(r,i)$ and $p(r,i)$. We find that achieving $6$-digit precision requires truncating the expansion at orders higher than $r^{8}$, $r^{12}$, $r^{18}$, and $r^{27}$ for $r = 0.01, 0.02, 0.03$ and $ 0.04$, respectively. This confirms the predicted convergence behavior in the relevant energy range.

\subsection{Higher-order $\alpha_s$ contribution}
To analyze the relative importance of each order in $\alpha_s$ expansion, we decompose the cross section as:
\begin{equation}
\begin{aligned}
 &\sigma \propto \left|\mathcal{A}^{\mu(0)}+\mathcal{A}^{\mu(1)}+ \mathcal{A}^{\mu(2)} + \mathcal{O}(\alpha_s^3)\right|^2 \\
 &=\mathcal{A}^{\mu(0)}\mathcal{A}_{\mu}^{(0)*} +2\operatorname{Re}\left[\mathcal{A}^{\mu(1)} \mathcal{A}_{\mu}^{\mu(0)*}\right] 
 +2\operatorname{Re}\left[\mathcal{A}^{\mu(2)} \mathcal{A}_{\mu}^{(0)*}\right] \\
 &+\mathcal{A}^{\mu(1)}\mathcal{A}_{\mu}^{(1)*} +2 \operatorname{Re}\left[\mathcal{A}^{\mu(2)} \mathcal{A}_{\mu}^{(1)*}\right] +\mathcal{A}^{\mu(2)}\mathcal{A}_{\mu}^{(2)*} \\
 &+\cdots ,
\end{aligned}
\end{equation}
where the terms are of order $\mathcal{O}(\alpha_s^0)$ (first term), $\mathcal{O}(\alpha_s^1)$ (second term), $\mathcal{O}(\alpha_s^2)$ (third and fourth terms), $\mathcal{O}(\alpha_s^3)$ (fifth term), and $\mathcal{O}(\alpha_s^4)$ (sixth terms), respectively.

\begin{figure}[htb]
    \begin{minipage}[b]{.9\linewidth}
        \centering
        \includegraphics[width=1\linewidth]{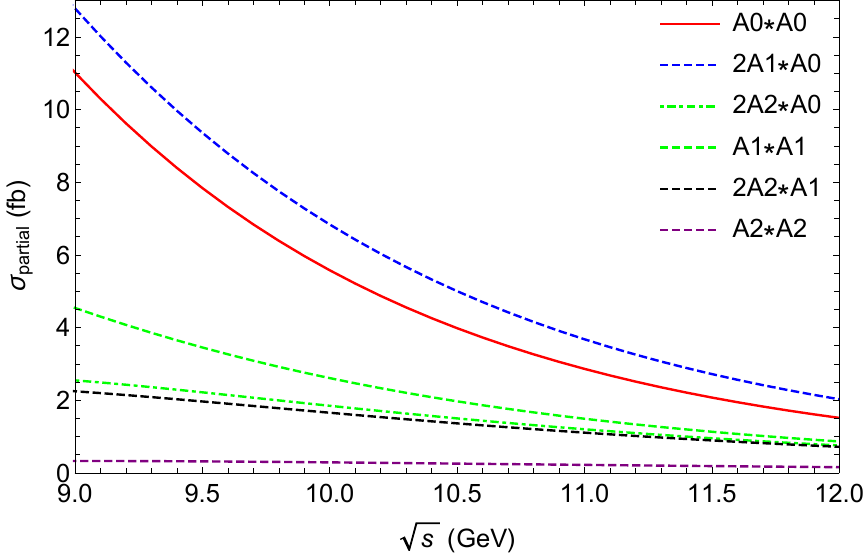}
    \subcaption{}\label{fig:ratioa}
    \end{minipage}
    \begin{minipage}[b]{.9\linewidth}
        \centering
        \includegraphics[width=1\linewidth]{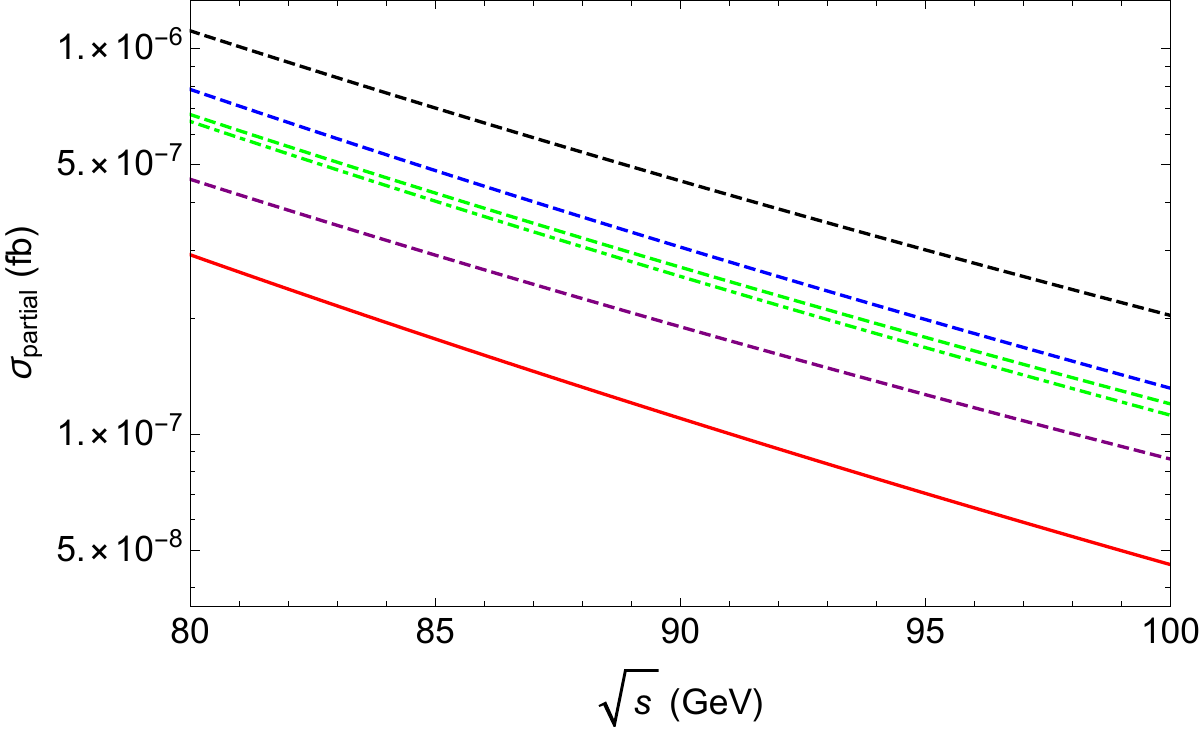}
    \subcaption{}\label{fig:ratiob}
    \end{minipage}
    \caption{The relative contribution of each term to the cross section of $e^{+} e^{-} \rightarrow \gamma^{\ast} \rightarrow J/\psi+\eta_{c}$ production at two different energy ranges: (a) 9 to 12~GeV and (b) 80 to 100~GeV. The renormalization scale is $\mu_R=\sqrt{s}$.}\label{fig:ratio}
\end{figure}

We plot the contribution of each term for $J/\psi+\eta_{c}$ production in Fig.~\ref{fig:ratio}. Both $\mathcal{O}(\alpha_s^2)$ components (two green dashed lines) are sizable compared to the LO contribution (red solid line). Furthermore, the $\mathcal{O}(\alpha_s^3)$ interference term, $2\operatorname{Re}[\mathcal{A}^{(2)} \mathcal{A}^{(1)*}]$ (black dashed line), remains significant, suggesting that the full $\mathcal{O}(\alpha_s^3)$ correction may yield non-negligible contributions.

Fig.~\ref{fig:ratioa} shows that near $\sqrt{s} = 10.6$ GeV, the partial $\mathcal{O}(\alpha_s^4)$ term (purple dashed line) is much smaller than all lower-order contributions. However, as shown in Fig.~\ref{fig:ratiob}, at higher energies this $\mathcal{O}(\alpha_s^4)$ term, while still smaller than the $\mathcal{O}(\alpha_s^1)$, $\mathcal{O}(\alpha_s^2)$, and $\mathcal{O}(\alpha_s^3)$ terms, becomes substantially larger than the $\mathcal{O}(\alpha_s^0)$ LO term. This behavior raises concerns about the reliability of fixed-order perturbative expansions and may call for resummation or alternative theoretical treatments.

\section{Summary and outlook}
In this work, we have computed the $e^{+} e^{-} \rightarrow J / \psi+\eta_c$ two-loop amplitudes within the framework of NRQCD. All divergences from different sources are systematically subtracted, yielding finite amplitudes which are presented both analytically and numerically as an expansion in the dimensionless variable $r=m_c^2/s$.
We provide a detailed analysis of the resulting cross section across various center-of-mass energies. By replacing $m_c$ by $m_b$ and changing the active number of flavors, we also extend our semi-analytical result to $\Upsilon+\eta_b$ production.

Our predictions are in good agreement with existing Belle and BaBar measurements, and we outline prospects for future experimental tests at Belle II and proposed high-energy lepton colliders. The analytical structure of our result provides a valuable check for the resummation formalism of large logarithmic terms in the future. Our results are presented at the amplitude level, and thus are well-suited for studying interference with other channels.

Nevertheless, certain theoretical uncertainties remain unsatisfactory. These include the dependence on the renormalization scale $\mu_R$, the convergence behavior of the perturbative QCD expansion, and the extraction of long-distance matrix elements  within the NRQCD formalism. These limitations highlight the need for continued theoretical and experimental efforts in precision quarkonium physics.

{\it Note added:}
While finalizing this paper, we noticed a paper \cite{Li:2025mng} presenting a similar calculation.
In this paper, not all the coefficients of the logarithms are expressed in analytic form.
%\wjcom{[Numerical agreement?]}

~\\

\begin{acknowledgments}
The authors would like to thank P. Zhang and X. Liu for useful discussion.
This work is supported in part by the National Natural Science Foundation of
China (Grants  No. 11875071, No. 11975029, No. 12075251, No. 12005117, No. 12321005, No. 12035007), the National Key Research and Development Program of China under
Contracts No. 2020YFA0406400, the High-performance Computing Platform of Peking University. XG is supported by the United States Department of Energy, Contract DE-AC02-76SF00515. XC is supported by the Swiss National Science Foundation (SNF) under contract 200020\_219367.
The Feynman diagrams in this paper are drawn with the aid of \texttt{FeynGame}~\cite{Bundgen:2025utt}.
\end{acknowledgments}

\bibliography{refs}

\end{document}